\documentstyle[12pt]{article}
\newcommand{\ncm}{\newcommand}
\ncm{\rencm}{\renewcommand}
\rencm{\thefootnote}{\mbox{\protect{$\fnsymbol{footnote}$}} }
\begin{document}

\begin{flushright}
BI-TP 94/52 \\
TIFR/TH/94-31 \\
hep-lat/9410004 \\
\end{flushright}
\bigskip
\bigskip

\begin{center}
\large{\bf  UNIVERSALITY AND THE DECONFINEMENT PHASE TRANSITION IN SU(2) LATTICE GAUGE THEORY } \\
\bigskip
\bigskip
\large {${\bf Manu Mathur^{(a)}}$\footnote {On leave from S. N. Bose National Centre 
for Basic Sciences, DB-17, Salt Lake, \\
\phantom{alcu}Calcutta-700 064, India.\\ 
\phantom{alcu}e-mail: mathur@theory.tifr.res.in} 
and  ${\bf Rajiv~ V.~ Gavai^{(a,b)}}$}\footnote{Alexander von Humboldt Fellow,
on leave from T. I. F. R., Bombay, India. \\
\phantom{alcu}e-mail: gavai@theory.tifr.res.in\\
\phantom{alcue-mail:} gavai@physw.uni-bielefeld.de} 
\bigskip

$^{(a)}$Theoretical Physics Group \\
Tata Institute of Fundamental Research \\
Homi Bhabha Road, Bombay 400 005, India \\
\bigskip

$^{(b)}$Fakult\"at f\"ur Physik\\
Universit\"at Bielefeld\\
Postfach 100 131\\
D-33501, Bielefeld, Germany\\
\end{center}
\bigskip
\bigskip
\begin{center}
\newpage
{\bf ABSTRACT}\\ \end{center}
\bigskip
\noindent  We study the three dimensional fundamental-adjoint $SU(2)$ 
lattice gauge theory at non-zero temperatures by Monte Carlo simulations.  
On an $8^3 \times 2$ lattice, at $\beta_A = 1.1$, where $\beta_A$ is the
adjoint coupling, we find no evidence of any transition at the location 
of a previously known bulk phase transition around $\beta = 1.33$.  
Moreover, the deconfinement transition at $ \beta_A = 1.1$ occurs at 
$\beta=1.20$ and is of first order for $\beta_A \ge 1.1$, thus 
implying a change of universality class from that of the Wilson action at 
$\beta_A=0$.  Computations of the plaquette susceptibility and the
temporal and spatial Polyakov loops on $8^3 
\times 4$ and $16^3 \times 8$ lattices at $\beta_A = 1.1$ further support these
conclusions and suggest that the previously claimed bulk 
transition around $\beta = 1.33$ is, in fact, the first order deconfinement 
transition.  Simulations at larger  $\beta_A$ and the measurements of 
the mass gaps from the correlation functions of temporal and spatial Polyakov 
loops also confirm the temperature dependent nature of the above transition. 
The consequences of our results on universality are discussed. 

\newpage

\begin{center}
\large {\bf 1. INTRODUCTION} \\
\end{center}
\bigskip

One of the most outstanding problem in physics is to understand the
mechanism for confinement and finite temperature deconfinement  in
quantum chromodynamics (QCD) or $SU(N)$ gauge theories in general.  
Colour confinement implies that the physical states in the Hilbert
space are colour singlets.  Non-abelian gauge theories at zero
temperature are believed to have this confining property.  Regulating
a compact gauge theory on the lattice with cut-off $a^{-1}$, where
$a$ is the lattice spacing, confinement can be analytically
established in the strong coupling region.  Creutz\cite{Cre} showed
that no phase transition takes place at zero temperature for the usual Wilson 
action of the $SU(2)$ or $SU(3)$ lattice gauge theory as the
continuum limit is approached, i.e., these theories have a
smooth $\beta$ function for all values of the cut-off, with a zero only when
the cut-off goes to infinity. 
In contrast, the  phase diagrams of lattice $SO(3)$, $SU(4)$ and $SU(5)$  
models have  bulk transitions separating the  corresponding strong and weak 
coupling phases.
However, Bhanot and Creutz\cite{BhaCre} showed that the mere existence 
of a bulk phase transition in lattice theories does not necessarily imply 
a loss of confinement.  As the lattice action of a given continuum 
theory is not unique, one could instead consider an extended lattice
theory belonging to the same 
universality class. If the phase diagram of the theory in the extended 
coupling space allows a 
smooth continuation around the bulk singularities then the theory will retain 
its strong coupling confining property as the cut-off is removed .   
They illustrated this by considering an extended Wilson action defined by:  
$$
S = \sum_P \beta\left[1 - {1 \over 2} Tr_F U_P\right] + \sum_P \beta_A\left[1 -
{1\over 3} Tr_A U_P\right]
\eqno (1)
$$
Here $F,A$ denote fundamental and adjoint representations
respectively, $U_P$ is the ordered product of the four directed link variables
$U_\mu(n) (\in SU(2))$ which form an elementary plaquette.  The sum over $P$ 
denotes the sum over all independent plaquettes of the lattice.
As found in Ref. \cite{BhaCre}, this model has a rich phase structure, 
which is displayed in Fig. 1.  At $\beta_A=0$, 
it reduces to the usual Wilson action. Along the $\beta=0$ axis, it
describes the SO(3) model with a first order phase transition at $\beta_A^c
\sim 2.5$. At $\beta_A=\infty$, it describes the $Z_2$ lattice gauge theory
with a first order phase transition at $\beta^c={1\over2} \ln(1+\sqrt{2}) \sim
0.44$. From the $\beta=0$ and $\beta_A=\infty$ axes, the above two bulk 
transitions extend into the $(\beta,\beta_A)$ plane, meet at a point ${\bf A}$ 
at (0.5,2.4) and then continue as another line of bulk first order phase 
transitions. The fact that the latter line ends at a point ${\bf B}$ 
at $(1.5,0.9)$ in the phase diagram allows one to bypass all the bulk 
singularities without losing confinement in $SU(2)$ gauge theory.
This model therefore retains its strong coupling 
confining property also in the continuum and is expected to be in the 
same universality class as the Wilson action $(\beta_A = 0)$.  

Comparing the naive classical continuum limit of Eq. (1) with
the continuum $SU(2)$ Yang Mills action, one obtains
$$
1/g^2_u = \beta/4 + 2\beta_A/3
\eqno (2)
$$
Here $g_u$ is the bare coupling constant of the continuum theory.  It
is convenient to characterise the model with $g_u$ and $\theta$, defined
by $\theta = \tan^{-1} {\beta_A / \beta}$.  The two loop $\beta$-function 
of this theory \cite{GonKor} tells us that theories characterised by 
different $\theta$'s approach the universal critical fixed point 
$g^{\star}_u = 0$ in the continuum limit,  with the flow governed by the
asymptotic scaling relation,
$$
a = 1/\Lambda(\theta) \exp\left[- {1 \over 2\beta_0 g^2_u}\right] \left
[\beta_0 ~g^2_u\right]^{-\beta_1/2\beta^2_0} ~~,~~
\eqno (3)
$$
where
$$
{\Lambda (0) \over \Lambda (\theta)} = {5\pi^2 \over 11} {6 \tan \theta
\over (3+8 \tan \theta)}.
\eqno (4)
$$

\noindent
Here $\beta_0$ and $\beta_1$ are the first two coefficients of $\beta$ function
for $SU(2)$ gauge theory.  Eqs. (3) and (4) show that the continuum limit of 
the theory lies at $\it {g_{u}^{\star}}$=0  and that $\theta$ is an irrelevant 
coupling which changes only the scale of the theory. Up to two loops this 
illustrates  that  the extended Wilson action described by Eq. (1) is in the 
same universality class as the Wilson action.

Usually finite temperature lattice field theory simulations are carried
out on $N_\sigma^3 \times N_\tau$ lattices, with $N_\sigma \gg N_\tau$
and the inverse physical temperature given by $ N_\tau a(g_u)$.  
Here $a(g_u)$ is the lattice spacing at the coupling $g_{u}$, given 
in the asymptotic scaling region by Eqs. (3-4).   The corresponding
physical volume is $V = N_\sigma^3 a^3$.  If $T_c$ denotes the
transition temperature of the deconfinement phase transition and is
thus also a physical observable, one gets the scaling relation 
$N_\tau \propto a^{-1}$ in the limit $ a \rightarrow 0$.  Together with
Eq. (3), this yields the scaling law for the critical coupling 
$g_u^c(N_\tau) $ for the lattice  with its temporal extent $N_\tau$.
If $N_\sigma$ is not very large compared to $N_\tau$, then one expects
corrections to this scaling relation which are, however, small: The
leading effect is still the exponential scaling of $N_\tau$ with the
corresponding critical coupling $g_u^c(N_\tau)$.  Furthermore, the relation 
between $\Lambda(\theta)$ and $\theta$, given in Eq. (4), is obtained in 
leading order perturbation theory.  Corrections to this relation are scheme 
dependent and not well known but they too must become vanishingly small for 
sufficiently large lattices.  Note that as the temporal lattice size becomes
infinitely large, i.e., as the continuum limit is approached, the 
critical coupling moves to zero: $g_u^{c}(N_\tau \rightarrow \infty) 
\rightarrow 0 $.  Thus in the language of finite size scaling theory
one can characterise the deconfinement phase transition as one with a
critical coupling $g_u^\star =0 $ and an effectively logarithmic scaling of
the critical coupling with the temporal lattice size $N_\tau$.

The finite size scaling behaviour of the shift of the critical coupling
(or temperature) of a bulk  transition \cite{Bar}, on the 
other hand, is governed by  
$$
[g^c_u (N) - g^c_u(\infty)] \sim N^{-\delta}  ~~,~~
\eqno (5a)
$$
and the height of the peak of the corresponding specific heat, which in 
our case is the plaquette susceptibility, should scale according 
to
$$ 
\chi_{max} \propto N^{\omega_s}~~.~~
\eqno (5b)
$$

\noindent
Here $\delta$ is the shift exponent, $N$ is the typical size of the
system : $N= (N_\sigma^3 N_\tau)^{1/4}$.  $g^c_u (N)$ and $g^c_u(\infty)$ 
are the critical couplings for the finite $N^4$ and the infinite lattices
respectively  and the location of $\chi_{max}$ can be used to define
the critical coupling, $g^c_u(N)$. For a second order bulk phase transition,
$\delta = 1 / \nu $ and $\omega_s={\alpha / \nu}$, where $\alpha$ and 
$\nu$ are the critical exponents of the plaquette susceptibility and the 
correlation length respectively. For a first order transition, $\omega$
is equal to the dimensionality of the lattice $d$ (=4, in our case). 
The bulk transitions in the phase diagram in Fig. 1 must exhibit 
scaling relations similar to Eq. (5-a,b) for various $\theta$ on the transition 
line {\bf AB}.  As the lattice size is increased, the bulk transition should move 
towards the non-zero $g_u^c(\theta)$, with a rapid  increase in the plaquette 
susceptibility, and remain anchored there as the lattice 
becomes infinitely large.  Thus unlike the finite temperature deconfinement 
transition, the bulk transition is a spurious transition not seen 
by the continuum limit of the theory at $g_u \rightarrow 0$. 

Recently we \cite{us} argued that the phase diagram in Fig. 1 is incomplete.
Since the diagram is largely based on Monte Carlo simulations on small
finite lattices $N_\sigma^4$ $(N_\sigma = 5$-7), it must necessarily
have a deconfinement phase transition at sufficiently small coupling
$g_u$.  Monitoring the order parameter $\langle L \rangle$
for the deconfinement phase transition on $N_\sigma^3 \times N_\tau$
lattices, for $N_\sigma=8$-12 and $N_\tau=4$ and 6, in order to understand
the phase diagram better, we found that
the known second order deconfinement transition at $\beta_A=0.0$
moves into the phase diagram as $\beta_A$ is increased and joins the first 
order bulk transition line around $\beta_A \simeq 1.0$ as shown by the dotted 
line ${\bf QR}$ in Fig. 1. This coincidence of the two transitions persists 
till  $\beta_A =1.5$ up to which the simulations were performed.  Moreover, 
there was no evidence for two separate, bulk and deconfining, transitions. 
The fluctuations in Polyakov loops and plaquettes were correlated and their 
discontinuities occurred at the same coupling.  Increasing the temporal size 
of the lattice resulted in a small shift in $\beta_c$ and the transitions still 
remained coincident and discontinuous.  While the critical exponents were 
in agreement with the Ising model, as predicted by universality\cite{SveYaf},
for $\beta_A \le 0.9$, there was strong evidence that the transition is
of first order for $\beta_A \ge 1.1$.  The histograms of $|L|$ showed 
a definite two peak structure with the dip between the two peaks increasing 
with $N_\sigma$ at $\beta_A=1.1$ and a co-existence of two phases with 
a tunnelling time $(> 10^5)$ at $\beta_A = 1.5$.

The above joining together of the deconfinement transition line with 
the previously claimed bulk transition line leads to a curious paradox 
because as discussed 
above, both their natures as well as  scaling behaviours are different. We 
therefore proposed the following three possible scenarios in our previous 
paper to explain the paradox :

\begin{enumerate}

\item[{A]}] The joining of the deconfinement transition line with the 
 bulk transition line is accidental on the lattices used. They will decouple 
 on larger (or smaller) lattices.
 
\item[{B]}] The entire transition line corresponds to deconfinement
 phase transitions only and the previous identification of it as a
 bulk transition line is, in fact, incorrect.
 
\item[{C]}] The whole transition line is due to bulk transitions,
 implying that the theory has no confinement in the continuum limit.

\end{enumerate}

In this paper we try to resolve the above paradox, at least 
partially, by studying the model on $8^3 \times 2$, $8^3 \times 4$ and $16^3 
\times 8$ lattices.  The spread in $N_\tau$ was thus maximised to
accentuate the scaling behaviour of the deconfinement phase transition.
On the $8^3\times 2$ lattice at $\beta_A=1.1$ we find a deconfining transition 
at $\beta=1.20$ with no evidence of any transition around $\beta=1.33$, where 
the bulk transition was expected. 
By also measuring the spatial Polyakov line $\langle L^\sigma \rangle$ 
in addition to the temporal Polyakov line $\langle L^\tau \rangle$, we 
attempted to increase the $N_\tau$ range further in the direction of asymptotic 
scaling region.  The simultaneous measurements of these on the
{\it same~lattice} have a major consequence for the bulk phase 
transition hypothesis; since the volume and the geometry of the lattice
is unchanged, one expects a bulk transition to occur at the same coupling 
$\beta$ for both of them. We find that the  transitions in $L^\tau$ and 
$L^\sigma$ are  located at {\it different} couplings for each of the lattices 
investigated. Although we do find the plaquette susceptibility to peak
in a range of $\beta$ close to the expected first order bulk transition
at $\beta_A=1.1$, the height of its peak is found to {\it decrease }
as the lattice volume increases.   
Further, the deconfinement phase transition at $\beta_A = 1.1$ appears to be 
a first order transition, whose strength increases rapidly as one
increases $\beta_A$:  clear coexistence of two-states was seen for
$\beta_A =1.4$ and 1.5.  The scaling behaviour of the transition with $N_
\tau$ as well as the computations of mass gaps 
obtained from our measurements of the correlation functions of the temporal 
and spatial Polyakov lines again confirm the deconfining and first order 
nature of the above transition.

The  organisation of the paper is as follows: In section 2 we describe
the observables used to monitor the phase transition.  Section 3
is devoted to the description of our Monte Carlo data and its analysis.  
Our results and their consequences on universality are discussed in the
next section.  Finally we conclude the paper with a summary and a 
listing of the still unanswered questions.

\begin{center}
\large{\bf 2.~~THE OBSERVABLES} \\
\end{center}
\bigskip

In order to understand the nature of the phase transitions around the 
transition points of the extended action, defined by
Eq. (1), we study the expectation values of the temporal 
and spatial Polyakov loops and the behaviour of their correlators in the 
two phases. As discussed below, these observables and their correlators 
characterise the phases on the two sides of the deconfinement transition. 
We also studied the expectation value of the plaquette, given by
$$
\langle P \rangle  = {1\over2} {\sum_p \langle Tr_F U_p  \rangle 
\over 6N^3_\sigma N_\tau}
\eqno (6)
$$
and the corresponding susceptibility,
$$
\chi_P = 6N^3_\sigma N_\tau (\langle P^2 \rangle - \langle P
\rangle^2)
\eqno (7)
$$
Here the expectation value of an observable $O$ is defined by
$$
\langle O \rangle = Z^{-1} \int \prod_{\mu, n} dU_\mu(n) O \exp
- S
\eqno (8)
$$
with $S$ as the extended action given by Eq. (1) and $Z$ as the
corresponding partition function.  

Due to periodic boundary conditions in the temporal direction, the
$SU(2)$ gauge theory  has a $Z(2)$ invariance corresponding to the 
center of the gauge group \cite{Sve}. Under this symmetry
$$
U_0 (\vec n,\tau_0) \rightarrow ~z~U_0 (\vec n,\tau_0),~~ \forall \vec
n~\&~{\rm fixed~\tau_0~~,~~ }
\eqno (9)
$$  
with $z \in Z(2)$.
Here $U_0 (\vec n,\tau)$ is the time like link at the lattice site
$(\vec n,\tau)$.  Under the above transformation, the Polyakov loop, 
which is the order parameter for the deconfinement phase transition and
is defined as
$$
L^\tau \equiv {1\over{N_\sigma^3}} \sum_{\vec n} L^\tau (\vec n)
= {1\over{2N_\sigma^3}} \sum_{\vec n}
Tr \prod^{N_\tau}_{\tau=1} U_0 (\vec n,\tau)~~,~~
\eqno (10)
$$
transforms as follows:
$$
L^\tau \rightarrow ~ z ~L^\tau.
\eqno (11)
$$
The expectation value of the Polyakov loop can be shown\cite{Sve,SveMac} 
to be a measure of the free energy of an isolated quark. Thus the 
deconfinement transition corresponds to spontaneous breaking of the $Z(2)$ 
symmetry, separating the $Z(2)$ symmetric low temperature (confined) phase  
from the high temperature broken (deconfined) phase.   Since due to
tunnellings between the two $Z(2)$-vacua the expectation value of this order 
parameter on a finite lattice is zero, we chose, as usual, to look at 
its absolute value and the corresponding susceptibility, defined by
$$
\chi = N^3_\sigma (\langle L^2_\tau \rangle - \langle |L^\tau| \rangle^2)
\eqno (12)
$$

On an infinite lattice the critical behaviours of the Polyakov loop, 
its susceptibility and the the correlation length of the theory 
are governed by the critical exponents $\beta,\gamma$ and $\nu $, defined 
as :

\begin{eqnarray}
~~~~~~~~~~~~~~~~~~~\langle L^\tau \rangle & \propto & |T - T_c|^\beta  {\rm ~~~~~for}
~{T \rightarrow T^+_c }~~~~~~~~~~~~~~~~~~~~~~~~~~~~(13) \nonumber\\[2mm]
\chi & \propto & |T - T_c|^{-\gamma} {\rm ~~~for}
~{T \rightarrow T_c }~~~~~~~~~~~~~~~~~~~~~~~~~~~~~~(14) \nonumber \\[2mm]
\xi & \propto &  |T - T_c|^{-\nu} {\rm ~~~for}
~{T \rightarrow T_c}~~~~~~~~~~~~~~~~~~~~~~~~~~~~~~(15) \nonumber
\label{critexp}
\end{eqnarray}
One obtains these exponents from Monte Carlo simulations by 
simply fitting the order parameter \cite{SveMac,EngFinWeb,GavSat} and/or
by using the finite size scaling theory\cite{Bar}.
According to the latter, the maximum of the susceptibility, 
defined in Eq. (12), is expected to grow like
$$ 
\chi_{\rm max} \propto N_\sigma^{\omega}~~,~~
\eqno (16)
$$
where $\omega=\gamma/\nu$ for a second order transition and is equal to 
spatial dimension of the system (3 in our case) for a first order transition.

In addition, we also define averaged spatial Polyakov loops by
$$
L^\sigma = {1 \over 3} \sum^3_{i=1} L^i_\sigma~~,~~
\eqno (17)
$$
with
$$
L^i_\sigma = {1 \over 2(N^2_\sigma N_\tau)} \sum_{n_0, \atop \vec n \bot \hat{i}}
Tr \prod^{N_\sigma}_{n_i =1} U_0 (\vec n,n_0)~~.~~
\eqno (18)
$$
The summation over $i $ is over the three
spatial directions and $\hat{i}$ are the corresponding unit vectors.  
Due to the periodic boundary conditions in the spatial 
directions also, $L^\sigma$ can again be thought of as an order parameter 
for the deconfinement transition now for a lattice of spatial volume
$N_\sigma^2 N_\tau$ and at an inverse temperature of $ N_\sigma a$.
Due to the larger effective ``temporal'' extent, the order parameter
$L^\sigma$ can be considered to probe a region closer to 
the asymptotic scaling region compared to that probed by $L^\tau$.  Of
course, the effective 3-volume, $N^2_\sigma N_\tau$,  seen by $L^\sigma$, being 
smaller than the corresponding spatial volume, $N^3_\sigma$, for $L^\tau$, 
finite size effects are relatively more in this case.  However, as discussed 
in the next section, on the lattices concerned these corrections only introduce
uncertainities of the order of a few per cent in the final results discussed in 
this paper.  Therefore, for our purpose of exploring the transition
region, spatial Polyakov loops are as good an order parameter as temporal 
Polyakov loops.  Moreover, due to the averaging over the three spatial 
directions, for $N_\tau > {N_\sigma\over 3}$, 
the statistics for $L^\sigma$ is also relatively better. 

The most important advantage of
measuring both the order parameter $L^\tau$ and $L^\sigma$ turns out to
be the ability to distinguish between the three scenarios mentioned
earlier.  Let us therefore consider them again in light of these
observables.

\begin{enumerate}

\item[{A]}] If both bulk and deconfining transitions are present and overlap
for the lattice sizes investigated in Ref. \cite{us},
then the effective increase in the temporal extent
by studying $L^\sigma$ may allow us to see the deconfinement phase
transition to shift away.  $L^\tau$ and
$L^\sigma$ should show critical behaviours at $\beta^\tau_c$ and $\beta^\sigma_
c$ respectively with $\beta^\sigma_c > \beta^\tau_c$.  At $\beta < \beta^\tau_c
$ both $L^\tau$ and $L^\sigma$ should show confinement.  In the region $\beta^
\tau_c < \beta < \beta^\sigma_c$, $L^\tau$ will be deconfined while $L^\sigma$ 
should still be in the confined phase.  For $\beta > \beta^\sigma_c$, both 
$L^\tau$ and $L^\sigma$ should show deconfinement.
A presence of a bulk phase transition may show up as a non-analyticity 
in $L^\sigma$ or in the corresponding mass gap obtained from 
$L^\sigma$-correlations to be defined below.

\item[{B]}] If there is only a deconfinement transition then 
the $L^\sigma$ or the corresponding mass gap will be smooth at
$\beta^\tau_c$. Moreover, in the asymptotic scaling region the difference 
$\beta_\sigma^c-\beta_\tau^c$ should decrease and approach a constant 
as we increase the lattice size holding $N_\sigma/N_\tau$ constant while
it should increase if the lattice size is increased only
in the  $N_\sigma$-directions.

\item[{C]}] If there is only a bulk transition present then $\beta^\tau_c 
\approx \beta^\tau_\sigma$ : both $L^\tau$ and $L^\sigma$ should become
significantly different from zero at about the same beta, especially
since they are evaluated on a lattice with identical 4-volume and
geometry.  Also the average plaquette should show a critical 
behaviour roughly at the same coupling. Unlike scenario [B],  the differences 
in the critical couplings should vanish in both the limits discussed
there.

\end{enumerate}

Another useful observable, mentioned already above, is the mass gap
obtained from the correlations of the Polyakov loops.  For this purpose,
we define sums of temporal and spatial Polyakov lines
over the corresponding orthogonal planes:
$$
{\cal L}^\tau(n_j) = {1 \over N^2_\sigma} \sum_{\vec n \bot \hat{j}}
L^\tau (\vec n)
\eqno (19)
$$
$$
{\cal L}^\sigma_{i} (n_j) = {1 \over N_\sigma N_\tau} \sum_{n_0,
\atop{\vec n \bot \hat{i}, \vec n \bot \hat{j}}} L^\sigma_i (n_0,\vec n)
\eqno (20)
$$
These can be thought of as the zero momentum projections of the Polyakov
lines in 2+1 dimensions.
In our simulations, we measure the zero momentum correlation functions 
of both the temporal and spatial Polyakov loops  given by:
$$
\Gamma^{\tau\tau(\sigma\sigma)}(|n|) = \langle{\cal L}^{\tau(\sigma)}(0)
{\cal L}^{\tau(\sigma)}(n) \rangle~~,~~
\eqno (21)
$$
Here  the indices on ${\cal L}^{(\tau)(\sigma)}$ have been  suppressed.  
Using periodic boundary conditions in the temporal and spatial directions 
and introducing the eigenstates of the transfer matrix as the intermediate 
set of states one obtains
$$
\Gamma^{\tau\tau(\sigma\sigma)}(|n|) = Z^{-1}\sum_{l,m}{\langle l|{\cal L}
^{\tau(\sigma)}(0)|m\rangle}^{2} \exp[-\mu_m^{\tau(\sigma)}|n|]
\exp[-\mu_l^{\tau(\sigma)}(N_\sigma-|n|)]~.
\eqno (22)
$$
In the large distance limit, ignoring the higher excited states of the transfer 
matrix and summing over only the lowest states, one can show that
$$
 \Gamma^{\tau\tau(\sigma\sigma)}(|n|) = {v^{\tau(\sigma)}}^2 \left\{\exp - 
 \mu^{\tau (\sigma)} |n| + \exp - \mu^{\tau(\sigma)}
 (N_\sigma-|n|]\right\} ~~,~~
 \eqno (23) 
 $$
with $ {v^{\tau(\sigma)}}^2 \equiv Z^{-1} {\langle 
0| {\cal L}^{\tau(\sigma)}|1 \rangle}^2 $.  $\mu^{\tau(\sigma)}$ are 
the mass gaps which can be shown\cite{GaKaPe,Jans,BerBil,EngMit} to correspond 
to the physical mass, namely, string tension, and the tunnelling mass in the 
confined and deconfined phases respectively.   This is because of the
expectation of a linearly rising colour averaged potential between a heavy 
quark and antiquark in the confined phase.  In the deconfined phase, the 
potential may become of the Debye-screened Coulomb form.  However, on a finite
lattice, the tunnelling between the two degenerate $Z(2)$-vacua,
$\mid 0_+ \rangle$ and $|0_{-} \rangle$  yields a symmetric $|0_{s} 
\rangle$ ground state and an antisymmetric first excited state 
$|0_{a} \rangle$, leading to a small ``tunnelling mass gap''. 
In the infinite volume limit, the tunnelling mass goes to zero and the
symmetry is spontaneously broken.   Thus this mass gap also acts as an
effective order parameter for the spontaneous breaking of the
$Z(2)$-symmetry and deconfinement.  Its main advantage over
the Polyakov loop is its direct relation with the string tension in the
confined phase.  Thus its becoming vanishingly small across a phase
transition is a clearer indication of it being a deconfinement phase 
transition.  Of course, the string tension will also decrease
across a bulk phase transition but its vanishing on the other side, 
especially as the lattice size grows to infinity, is very unlikely.

\begin{center}
\large {\bf 3.~~DATA AND ANALYSIS} \\
\end{center}
\bigskip

Our Monte Carlo simulations were done mostly at $\beta_A=1.1$ on 
$8^3 \times 2$, $8^3 \times 4$ and $16^3 \times 8$ lattices, using 
the Metropolis  {\it et~al.} algorithm.  On the $8^3 \times 2$ lattice we
also performed simulations at larger $\beta_A$ to confirm the order of
the phase transition and to estimate the shift in critical $\beta$.
Additional simulations were also made on $10^3 \times 2$ and 
$12^3 \times 2$ lattices to look for the finite size scaling behaviour and
the critical exponent $\omega$, defined in section 2.  Typically 1-2
$\times 10^5$ iterations were done on all lattices with about a factor
two less statistics on the biggest lattice.  For the  $8^3 \times 4$ 
and $16^3 \times 8$ lattices, we also measured the spatial Polyakov loops.  
Measurements of temporal and spatial zero  momentum correlators were 
done on the $16^3 \times 8$ lattice both in the confined and deconfined 
regions.  The distribution  function of the Polyakov loop is a good tool
for identifying  the critical points to a reasonably good accuracy, especially 
for a first order transition on larger lattices.  In the confined phase
one expects this function to be a gaussian peaked around zero, while in
the deconfined phase of the SU(2) gauge theory one expects two symmetric
gaussian peaks located away from zero.  At the transition point of a first 
order phase transition, one would have an additional peak of the same
height located at zero.  One can alternatively look at the 
order parameter, $|L|$, and correspondingly obtain a two peak structure. 
Even when the transition is of second order, or the precise location of the
transition point is not found, these distribution functions can be used
to set the lower and upper limits on $\beta_c$.  For our
semi-quantitative purposes of distinguishing the shifts due to the two
different scaling behaviours of the deconfinement and the bulk phase
transition on $8^3 \times 4$ and $16^3 \times 8 $ lattices, it was usually 
enough to study the behaviour of the histograms of the temporal and spatial 
Polyakov loops at various $\beta$-values.

\newpage
\begin{center}
{\bf 3.1~~$\bf N_\tau=2$} \\
\end{center}
\bigskip

Variation of the temporal size of the lattice should shift the deconfinement 
phase transition much more than the bulk phase transition, allowing one
to test the scenario [A] of the previous section.  Already in Ref.~\cite{us},
a change from $N_\tau=4$ to 6 was attempted but the results were
inconclusive as the shift in the critical coupling
was too small and only one transition was
observed.  From the scaling law for the deconfinement phase transition,
one expects a much larger shift by going to {\it smaller} $N_\tau$,
although it usually takes one in the strong coupling region.
Since our aim here is only to see whether there are two separate
transitions, it is perhaps not so important that one does not work in the
scaling region; hopefully, such qualitative aspects are not altered by
the continuum limit.  

Figs. 2a and 2c show the average temporal Polyakov
line $ \langle L_\tau \rangle $ and the average plaquette as a function
of the coupling $\beta$ for $\beta_A = 1.1$ on an $8^3 \times 2$ lattice.
The later value was chosen because our earlier work\cite{us}
suggests a weak first order phase transition at $(\beta, \beta_A) =$ 
(1.33, 1.1) on an $ 8^3 \times 4 $ lattice which could be coincident
with the bulk transition seen in Ref.~\cite{BhaCre} .  The
corresponding susceptibilities are displayed in Figs. 2b and 2d.  
Both the Polyakov line and the plaquette have a sharp and large jump 
around $\beta=1.20$ and correspondingly the respective susceptibilities 
exhibit a sharp peak there.  The above 
behaviour of the Polyakov line and its susceptibility clearly indicates 
the presence of a deconfinement transition at $\beta_c=1.20$. 
Note, however, that the behaviour of the Polyakov line and the plaquette 
is smooth in the region around $\beta=1.33$ where the bulk transition 
was expected.   For a better comparison a magnified view of the Polyakov line 
and the plaquette in the regions around the deconfinement phase
transition and around $\beta=1.33$ is shown in the Figs. 3a - 3d. In
order to show a bigger region around $\beta=1.33$, ranges of both x and y axis
have been enlarged by a factor of four in Figs. 3b and 3d as 
compared to those of Figs. 3a and 3c.  The remarkable flatness of the
data in Figs. 3b and 3d, together with the absence of any structure in the
susceptibilities in Figs. 2b and 2d, make it clear that there  
is no trace of any transition in the region around $\beta=1.33$. 
In view of the different scaling laws for the deconfinement and bulk phase
transitions and the additional results from Ref.~\cite{us}, indicating
a very small shift in $\beta_c$ in going from an $8^3 \times 4$ lattice
to a $12^3 \times 6 $ lattice, this relatively large shift makes it
difficult to sustain the accidental coincidence hypothesis in scenario [A].
Of course, the lattices used may not be large enough for various corrections
to the scaling behaviour to be small.  Nevertheless, it would be
remarkable, if not impossible, if they are similar in magnitude and 
sign for the two
unrelated scaling forms.  

The large shift in the critical coupling in going to $N_\tau=2$ from $N_\tau=
4$ and 6, and the fact that the order parameter $ \langle | L | \rangle$
shows a spontaneous breakdown of the $Z(2)$ symmetry at the transition
point are indicative of the deconfining nature of the phase transition.
Moreover, the histograms of $L$, on $8^3 \times 2$, $10^3 \times 2$ 
and $12^3 \times 2$ show a three peak structure at the transition point 
which, however, is not very pronounced at $\beta_A=1.1$.  
Correspondingly, the determination of the critical exponent $\omega$ 
does not fix the order of the phase transition uniquely. 
This is similar to the $N_\tau=4$ results in Ref.~\cite{us}.
It was found there that the above transition 
was a relatively weak first order transition at $\beta_A=1.1$ but for
$\beta_A > 1.1$ the discontinuity increased with an increasing 
$\beta_A$.   Having thus verified that the order of the phase transition 
does not change as $N_\tau$ is changed to 2 from 4, 
we simulated the theory at $\beta_A=1.4$ and 1.5 on the above 
lattices with $N_\tau=2$ to see if its strength increases.  
It does indeed increase and we were even able to confirm the first order
nature by determining the critical exponent $\omega$.
The evolution curves for both the plaquette and the Polyakov loop in 
the respective critical regions at $\beta_A=1.4$ are plotted in the Figs.
4a-4c and the corresponding histograms for the latter are shown in Fig. 5.  
One sees that they all suggest a clear first order deconfinement phase 
transition.  One does observe a sharpening of the transition as the volume is
increased.  It may be interesting to note that the average plaquette
also shows a discontinuity and fluctuations which are totally correlated
to that of $L$, which is again in agreement with the similar observation
on $N_\tau=4$ lattice.  
Individual runs around the critical coupling were used to locate the maximum 
of the Polyakov susceptibility. This maximum value of $\chi$ was in turn used 
to compute the exponent $\omega$, defined earlier by Eq.~(16). 
On the $12^3 \times 2$ lattice, a large tunnelling time  of 
$( \approx $30,000-40,000 sweeps) was observed at $\beta_A=1.4$, and
hence a relatively higher statistics run of $3 \times 10^5$ was made at
the transition point to ensure a better sampling of both the phases.  
The critical couplings and the estimated values
of the exponent $\omega$ are given in the Table 1. They are in excellent 
agreement with the finite size scaling prediction for a first order
deconfinement phase transition.  Finally, also the simulations at
$\beta_A=1.5$ support the above picture and clear three-peak histograms
are observed.  The critical coupling for the $ 8^3 \times 2$ lattice
for $\beta_A =$1.5
is $\beta=0.983 \pm 0.001$.  Comparing this value with the corresponding
$N_\tau=4$ result, one sees that the shifts due to a change of $N_\tau$
from 4 to 2 become smaller 
as $\beta_A$ grows larger, but even at $\beta_A=1.5$ the shift is 
nevertheless significantly large enough to be sure that it is
non-zero.  The discontinuity in $|L|$ also appears to increase slightly
on the smaller $N_\tau$ lattice when compared at the same $\beta_A$.
 
\begin{center}
{\bf 3.2~~${\bf N_\tau=4}$ and 8 } \\
\end{center}
\bigskip

The above results on the $N_\tau=2$ lattices showed clearly the absence of
any transition at the location of a previously claimed bulk phase
transition and further showed a first order phase transition at a much
smaller $\beta$.  The latter can be consistently interpreted as a
deconfinement phase transition.  While they make the coincidence of a
bulk transition unlikely, one would like to establish it as concretely
as possible.  With such a motivation,
we simulated the extended action at $\beta_A=1.1$ on $8^3
\times 4$ and $ 16^3 \times 8$ lattices.
Now, we also monitored the $L^\sigma$ since the corresponding effective 
3-volume is substantial on these relatively bigger lattices.

The histograms for $L^\tau$ and $L^\sigma$ on the $8^3 \times 4$ lattice 
at four values of the coupling $\beta$ = 1.31, 1.32635, 1.4 and 1.65 are shown 
in Figs. 6a-6d. These couplings have been chosen such
that the two intermediate couplings are close to the corresponding
$\beta^\tau_c$ and $\beta^\sigma_c$.  
Due to the weakness of the first order transition,
it was not possible to get a clear three-peak structure of the histogram.  
However, the reasonably flat behaviours of the $L_\tau$ and $L_\sigma$ 
in the Figs. 6b and 6c show the onset of the deconfining transition 
and suggest that it is first order.   Locating the critical point by the 
criterion of maximum susceptibility, $\beta^\tau_c$ was found in 
Ref.~\cite{us} to be 1.327 which is good agreement with the  
$\beta^\tau_c =1.32635$ of Fig. 6c. 
Figs. 6a-6d, and our additional simulations (not shown in the these figures) 
just below and above the critical couplings, show that these distributions 
follow the expected pattern of the ${\it deconfinement~transition}$ very 
neatly:

\noindent a] For $\beta < \beta^\tau_c$, both $L_\tau$ 
and $L_\sigma$ are confined,\\
b] at $\beta =\beta_\tau^c=1.32635$, $L_\tau$ is critical and $L_\sigma$ is 
confined,\\ 
c] at $\beta=\beta_\sigma^c =1.4, L_\sigma$ is critical and $L_\tau$ 
is deconfined,\\ 
d] for $\beta > \beta_\sigma^c$, both $L_\tau$ and $L_\sigma$ are 
deconfined. 

A similarly chosen pattern of histograms on the $16^3 \times 8$ lattice 
at five values of the coupling $\beta$=1.33, 1.35, 1.3508, 1.47 and 
1.57 is shown in Figs. 7a-7d.  Again, we observe a confinement-deconfinement 
pattern in $L_\tau$ and $L_\sigma$ like the one discussed above.
Due to the narrower transition region, we were unable to get 
better histograms of $L_\tau$ and fix the critical coupling $\beta_\tau^c$ 
more accurately.  However, as is clear from the Fig. 6b, the critical coupling
$\beta_\tau^c(N_\tau=8)$ is between 1.35 and 1.3508.   We have also observed a
clear signal for the coexistence of two phases at $\beta=1.35$,
with a discontinuity in the Polyakov line $L_\tau$ which is
roughly 0.05.  The corresponding value at the critical point on 
the $ 8^3 \times 4$ lattice was found to be 0.25.  Due to the
fluctuations in plaquette, its discontinuity could not be estimated but
it appeared to be smaller than the one observed on the $8^3 \times 4$
lattice as well.  Such a scaling of discontinuities is expected for a
deconfinement transition but not a bulk transition.  
Fig. 8a illustrates the scaling of the Polyakov line 
and the deconfinement temperature at $\beta_A=1.1$ as we 
increase the temporal lattice size and thus approach towards the continuum
limit. One can clearly see  the transition moving 
towards the continuum critical fixed point $g_u^{\star}$=0 with a
corresponding reduction in the discontinuity of $L^\tau$. On the other 
hand the scaling behaviour of the plaquette susceptibility at $\beta_A=1.1$
on the same lattices, namely on $8^3\times 2$, $8^3\times 4$ and $16^3\times 8$ 
lattices, shown in Fig. 8b, does not show any bulk phase transition.
It should be noted that on the largest lattice the peak of the
susceptibility is indeed at a bit lower value of $\beta$ compared to the
location of the deconfinement transition hinted in Figs. 7-b and 8-a.  
Such a behaviour would be consistent with the scenario [A] where one expects the
deconfinement phase transition to move away faster.  However, the height
of the peak of the susceptibility 
for the $16^3 \times 8$ lattice is seen to decrease in Fig. 8b
although the bulk volume increases by a factor of 16.  Combined with the
expected finite size scaling behaviour of a bulk first order  
in Eq. (5b), this result clearly rules out the presence of bulk transition 
convincingly in this region too. 

Comparing the critical value $\beta^\sigma_c=1.4$ for $L^\sigma \equiv L(8)$ 
on the $8^3 \times 4$ with the more precise value $\beta^\tau_c = 1.3504 
\pm 0.0004$ for $L^\tau \equiv L(8)$ on the $16^3 \times 8$ lattice, one
finds that the former is slightly overestimated .  If one uses the
maximum of the susceptibility $\chi(|L_8|)$ to define the transition
point on the $8^3 \times 4$ lattice on the other hand,
then one obtains $\beta_c^\sigma$=1.348, where the
Ferrenberg-Swendsen\cite{FerSwe} histogramming method was used on the 
configurations generated at $\beta$ =1.4 to locate the peak.   This 
relatively large difference in the determination of $\beta_c$ by two 
different methods for the $8^3 \times 4$ lattice is perhaps indicative 
of the smallness of the effective 3-volume for $L_8$.   Even on this
small lattice thus the errors in determination of $\beta_c$ by employing 
$L_\sigma$ as the order parameter are $\approx 5$ \% and a quantitative
estimate of $\beta_c$ on an effectively larger temporal lattice 
is thus feasible.  Since we are anyway only interested in seeing whether 
the transition indeed marches towards $g_u^{\star}=0$, such small 
shifts are perhaps tolerable to us, considering that it will
allow us to probe $N_\tau=16$ by studying $L^\sigma$ on $16^3 \times 8$
lattice.  With this caveat in mind, we see from Fig. 7c that $\beta_c
\simeq 1.47$ for $N_\tau=16$, as the  $L_{16}$-distribution can be seen 
to become very flat at $L =0$ on the $16^3 \times 8$ lattice at this
coupling.  Furthermore, using the configurations at $\beta=1.47$ to locate 
the peak of the $|L|$-susceptibility, one obtains $\beta^\sigma_c$ = 1.468.
Considering the increased effective 3-volume in this case, such an
excellent agreement in these two determinations is in accord with the
expectations.  Allowing for a few per cent overestimation nevertheless
due to the unusual geometry, one still sees that the transition is indeed slowly
moving towards $g_u=0$, although compared to the $\beta_A=0.0$ results
of Refs.~\cite{EngFinMil,FinHelKar} these results are still far away from 
exhibiting a consistency with the two-loop scaling equations in Eq. (3-4).

The systematic behaviour of the histograms in both Figs. 6 and 7, 
suggesting that $L^\tau$ and $L^\sigma$ start being non-zero at {\it different}
locations with the former doing so always earlier, has important
consequences for the scenarios [A] and [C] of sect. 2.  Since the
4-volume for each figure is held constant, and so is the geometry of the
lattice, merely one bulk phase transition would be inadequate to explain
this behaviour of the Monte Carlo data.  Thus, the scenario [C] is
untenable.   The strong correlation of the critical coupling with
only the linear extent of the lattice along which the order parameter 
is defined, taken together with the decrease in the plaquette
susceptibility at its peak in Fig. 8b and the results of previous subsection 
for $N_\tau=2$ lattice, is also inconsistent with scenario [A], thus leaving
the scenario B of a pure deconfinement phase transition as the only
plausible one.  Furthermore, the above mentioned decrease in the size of
discontinuities in the lattice observables, namely the Polyakov line and
the plaquette, again confirms the confining-deconfining nature of the 
transition.  For a bulk transition these discontinuities would be expected 
to remain unaltered or increase, as the total volume of the lattice increases. 

\newpage
\begin{center}
{\bf 3.3~~Polyakov Loop Correlations } \\
\end{center}
\bigskip

As a final check of the nature of the transition along the line ${\bf QR}$,
which shifts strongly with the ``temporal'' extent of the lattice, as we 
saw above, we have investigated the behaviour of the Polyakov loop
correlations and the corresponding mass gaps on the $16^3 \times 8$
lattice.  From the discussion above and in sect. 2, one expects to see
the mass gaps from the temporal and spatial Polyakov loop correlations 
to decrease strongly at the respective critical couplings.  Moreover,
they should roughly be the same, apart from finite size effects, in the
phases where both the order parameters show the same behaviour.
 
The measurements of the zero momentum projected temporal and spatial 
correlation functions both above and below 
the transitions were done by recording their values after every 30 sweeps.
The individual averages of temporal and spatial correlations were 
taken by summing $\Gamma^{\tau\tau}(n_j)$ over the three directions 
(j) and $\Gamma_{i}^{\sigma\sigma}(n_{j})$ over (i,j) for fixed 
value of $n_{j}$. The discrete rotational covariance 
was ensured by plotting them for each of the three spatial directions 
(i=1,2,3) individually and ensuring that their differences lie within 
error bars. The above averaged values of the temporal and spatial correlators 
are plotted in the Figs. 9a-9b.  A gradual flattening of the correlation
functions is clearly seen in both Figs. 9a and 9b, as one increases the
coupling $\beta$ through the respective $\beta_c$.  Below the transition
they all fall rather steeply while a very flat behaviour is evident for
very large $\beta$.  The  mass gaps in the confined and deconfined phases 
were obtained  by fitting the correlation function with a single 
hyperbolic cosine function. Their values and the corresponding errors
are given in Table 2 and are plotted in Fig. 10.  The relatively large
fluctuations suggest that much higher statistics may be needed for a
precise and reliable determination of these mass gaps.  Nevertheless,
the small value of the mass 
gap above the respective phase transitions, indicated by the corresponding
Polyakov line, is a clear signal of the $Z(2)$ symmetry 
being broken spontaneously in the infinite volume limit, leading to 
deconfinement of quarks.  The relatively larger finite size effects for
the spatial correlations are evident in the slower fall-off of the
corresponding mass gaps, especially  in the deconfinement region.
The results are again in agreement with the
conclusions of the previous subsections: one has deconfinement phase
transitions at the respective $\beta_c$.  Although the data are 
somewhat noisy and thus not very
conclusive, there is again no evidence of any extra first order bulk phase
transition.  It would have been nicer to have the spatial mass gaps at
still lower $\beta$ to confirm a continuity in them.  However, we found
a very rapid deterioration of the signal in the correlation functions in
that region and large fluctuations in the correlation functions
prevented us from arriving at any values.

\newpage

\begin{center}
\large {\bf 4.~~DISCUSSION AND CONCLUSIONS} \\
\end{center}
\bigskip

Our results in the previous section strongly suggest that the previously
identified line of first order bulk phase transitions, shown in Fig. 1 by
the line ${\bf QR}$, is in fact a line of first order deconfinement phase
transitions.  No bulk phase transition seems to be present in that
region.  Since this bulk transition and its ``shadow'' on the
$\beta_A=0$ axis have been speculated to have an impact on the onset of
confinement in the $SU(2)$ gauge theory at zero temperature, it is
clearly necessary to investigate the phase diagram of Fig. 1 afresh on
much bigger symmetric lattices than used in the past.  Similar structure
is also known to exist for other $SU(N)$ gauge theories, in particular
for the $SU(3)$ theory as well.  These may need to be investigated
further to investigate their differences.  

A first order deconfinement phase transition for $\beta_A \ge 1.1$ also
implies a change of the universality class for the extended action 
at and above $\beta_A=1.1$. It has been argued \cite{SveYaf}
that by integrating out the irrelevant
degrees of freedom of the $SU(2)$ theory at finite temperature, one
can obtain an effective theory of the order parameter $L$.  If this
theory and the Ising model in 3-dimensions, which too has a global
$Z(2)$ symmetry for its order parameter, have only one fixed point in
the space of couplings, then the two are in the same universality 
class.  This leads to a prediction of various critical exponents for the
former.  For $\beta_A = 0$, these universality predictions were explicitly 
verified by high precision Monte Carlo simulations of Engels et 
al.\cite{EngFinWeb}. They found ${\beta \over \nu} =0.545 \pm
0.030$, $\omega={\gamma \over \nu} = 1.93 \pm 0.03$, $\nu=0.65 \pm 0.04$. 
The corresponding values for the Ising model are ${\beta \over \nu} =0.516 
\pm 0.005$, ${\gamma \over \nu} = 1.965 \pm 0.005$, $\nu=0.63 \pm 0.003$. 
The change of the order of the transition 
(e.g at $\beta_A=1.4, \omega \approx 3$) 
as a function of an apparently irrelevant coupling $\beta_A$ shows a 
non-universal behaviour of the deconfining transition, which needs to
be understood in the renormalisation group picture.  It does cast a doubt
on our understanding of the order of the deconfinement phase transitions
from those in the corresponding spin models.  Of course, one cannot rule
out the possibility that this change of universality class is a purely
strong coupling phenomenon; simulations on sufficiently larger lattices
will show a second order deconfinement phase transition even at the
large $\beta_A$ used in this work.  It would be very interesting, and
illuminating, to check this since no such instance of a change of the order
of the phase transition in going over to the scaling regime is
known so far.  If confirmed, it would underline the importance
of checking even such qualitative aspects of full QCD thermodynamics
in the scaling or the asymptotic scaling regime.

A non-universal behaviour of the string tension depending on the lattice 
action used has been a question of debate in the past. In the SU(2) lattice 
gauge theory the $\Lambda$-ratios of the Manton and the  Heat kernel action 
with that of the Wilson action were found to differ from their theoretically 
predicted values. However, these discrepancies were shown\cite{GavKarSat} 
to be due to higher order corrections to the theoretical values which
were obtained in leading order perturbation theory. By taking the ratios 
of the physical observables in Monte Carlo simulations or by considering 
the effects of higher order terms the above actions were shown 
to be in the same universality class.  In the case of the extended SU(2) 
action of Eq.~(1), Bhanot and Dashen \cite{BhaDas} found that at $\beta_A=0.9$,
the $\Lambda$-ratio, $\Lambda _{Extended}/\Lambda_ {Wilson}$, obtained from
the computations of string tension in Monte Carlo simulations, was roughly 
a factor of four higher than its perturbation theoretical value in
leading order.   Again, this was shown\cite{GavKarSat,GonKorPeiPer} to
be due to the strong coupling effects. In fact, the string 
tension data compared well with its strong coupling expansion \cite
{GonKorPeiPer} and correspondingly the 3-loop contribution to $\Lambda$ 
ratio was quite large \cite{GavKarSat}.   

What we find in the change of the order of the deconfinement phase
transition, however, is a loss of universality in qualitative features.
and therefore should be considered much more seriously. 
Moreover, the line of deconfinement transition we investigated on $N_\tau
=4$ is remarkably close to the line of constant string tension, obtained
from Monte Carlo simulation \cite{BhaDas}, at $\sigma a^2=0.14$. This line 
lies in the cross-over region\cite{GonKorPeiPer}. Thus our results and 
conclusions could possibly be free from any strong coupling artifacts.
In other words, one need not expect that by going to larger temporal lattices 
at $\beta_A > 1.0$ the scaling exponent $\omega$ will drop from $3$ to its 
second order value $1.97$, corresponding to the 3-dimensional Ising model.   
Taking seriously the above coincidence of the lines of the
deconfinement transition and the constant string tension,
along with the fact that up to 2-loops of the $\beta$ function 
physics is independent of the value of $\theta$, the above 
change of universality class indicates a possibility  of  
a new fixed point and  some non-perturbative effects.  
However, much more work is necessary to convincingly establish or 
rule out the above interesting possibilities.

In the case of pure Wilson action, Polonyi and Szlachanyi \cite{PolSzl}
showed the second order nature of the deconfining transition
by computing  the effective action in terms of the order parameter $L$ 
in the strong coupling limit. It will be interesting to compute the effect 
of adjoint coupling $\beta_A$ and see how it changes the order of the 
transition. One expects a $L^4$ term with its coefficient becoming 
negative  as the adjoint coupling increases above $\beta_A=1.1$. 

Let us summarise by stressing that the universality hypothesis in lattice 
gauge theories, in particular of extended Wilson action,  should be 
be looked at more critically. It will be interesting to study similar 
phenomenon with the corresponding extended SU(3) Wilson action also.  

\begin{center}
\large {\bf 5. ACKNOWLEDGEMENTS} \\
\end{center}

It is a pleasure to thank Prof. Michael Grady, SUNY, Fredonia, USA for the
inspiring discussions which lead to this work.  Part of this work was
done when one of us (RVG) visited the University of Bielefeld as an
Alexander von Humboldt Fellow.  The financial support from the Humboldt
Foundation and the hospitality of the Physics Department in Bielefeld
are gratefully acknowledged.  We also thank Mr. V. S. N. Reddy and the 
staff of the computer center at Tata Institute of Fundamental Research,
Bombay for their help and for letting us use the ALPHA machines while 
they were being installed.

\newpage
\begin{center}
\bf{FIGURE CAPTIONS}
\end{center}

\bigskip

\noindent Fig.1 
The phase diagram of the extended SU(2) lattice gauge theory.  
The solid lines are from simulations done on a $5^4$ lattice by 
Bhanot and Creutz\cite{BhaDas}.  The broken lines ${\bf QR}$  
is  the finite temperature deconfinement phase transition line
on the $8^3 \times 4$ lattice. The deconfinement line  ${\bf ST}$ 
corresponds to $8^3 \times 2$ lattice. 
\bigskip

\noindent Fig.2
(a)Average temporal Polyakov loop, (b) its susceptibility, (c) 
average plaquette, and (d) its susceptibility vs. $\beta$ on 
an $8^3 \times 2$ lattice at $\beta_A=1.1$.
\bigskip

\noindent Fig.3
Average temporal Polyakov loop and average plaquette as a function of
$\beta$ on an $8^3 \times 2$ lattice in the regions of the deconfinement
phase transition (a, c) and the expected bulk transition (b, d).
\bigskip

\noindent Fig.4
Evolution of $L_\tau$ and (P+0.35) $\beta_A=1.4$ and on (a) $8^3 \times 2$ 
($\beta_c=1.03954$) (b) $10^3 \times 2$ ($\beta_c=1.0398$) and 
(c) $12^3 \times 2$ ($\beta_c=1.04007$)
\bigskip

\noindent Fig.5
Normalised probability distributions of $L_\tau$  on  an $8^3 \times 2$, 
$10^3 \times 2$ and  $12^3 \times 2$ lattices at $\beta_A=1.4$ and  
$\beta_c$=1.03954, 1.0398 and 1.04007 respectively. 
\bigskip

\noindent Fig. 6 
Normalised probability distributions of $L^\tau$ and $L^\sigma$ on $8^3\times 
4$ lattice at $\beta_A=1.1$ and (a) $\beta=1.31$, 
(b) $ \beta=\beta^\tau_c=1.32635$, 
(c) $\beta=\beta ^\sigma_c \approx 1.4$ and (d) $\beta=1.65$
\bigskip

\noindent Fig. 7 
Same as Fig. 6 but for a $16^3\times 8$ lattice and
at (a) $\beta=1.33$, (b) $ \beta=\beta^\tau_c=1.35,1.3508$, 
(c) $\beta=\beta^\sigma_c \approx 1.47$ and (d) $\beta=1.57$
\bigskip

\noindent Fig. 8
The nature of the first order transition at $\beta_A=1.1$: 
(a) The scaling behaviour of $L^\tau$ on $8^3 \times 2$, $8^3 \times 4$  
and  $16^3 \times 8$ lattices in accordance with the deconfinement 
transition and (b) the scaling behaviour of the plaquette susceptibility, 
ruling out any bulk transition in this region. 
\bigskip

\noindent Fig. 9
The (a) temporal and (b) spatial correlation functions on $16^3\times 8$ 
lattices at $\beta_A=1.1$ as a function of distance for various
couplings $\beta$ through the respective $\beta_c$.
\bigskip

\noindent Fig. 10 
The temporal and spatial mass gaps $\mu^\tau$ and $\mu^\sigma$ on $16^3\times 
8$ lattice at $\beta_A=1.1$.

\newpage 
\pagestyle{empty}
\begin{table}
\begin{center}
{Table 1}
\end{center}
The values of the critical couplings ($\beta_c,\beta_A=1.4$) on $N_\tau$=2 lattices, the \\
 $L^\tau$ finite size scaling exponent $\omega$.  The expected value for $\omega$ is 3.0(1.97)\\
 if the deconfining phase transition is first order (second order). \\ \\
\medskip
\begin{tabular}{|c|c|c|c|}
\hline
                         &                         &                       \\
~~$N_\sigma~~~~~~~~$ & $~~~~~~~\beta_c~~~~~~~$ & $ ~~~~~~~~~~~~~~~~~~~~~~~~\omega~~~~~~~~~~~~~~~~~~~~~~~~~~$ \\
                         &                         &                              \\
\hline \hline
    &      &                              \\
8 &  1.039540    &              ----              \\
    &      &                                 \\
\hline
    &      &                                 \\
10 & 1.03980    &   3.246(243) [$ 8^3 \times 2: 10^3 \times 2$]                 \\
    &      &                                 \\
\hline
    &      &                                 \\
12  &1.040070   &  3.204(184) [$12^3 \times 2 : 8^3 \times 2$]\\  
   &            &   3.154(406) [$10^3 \times 2: 12^3 \times 2$]                 \\
     &       &                                  \\
\hline
\end{tabular} 
\end{table}

\newpage
\pagestyle{empty}
\begin{table}
\begin{center}
{Table 2}
\end{center}
The values of the temporal and spatial mass gaps, $\mu^\tau$ and $\mu^\sigma$,   at $\beta_A=1.1$ on $16^3 \times 8$ lattice. \\ \\ 
\medskip
\begin{tabular}{|c|c|c|c|}
\hline
                         &                         &                       \\
~~~~$\beta~~~~~~~~$ & $~~~~~~~~~~~~~~~~~\mu^\tau~~~~~~~~~~~~~~~~~$ & $ ~~~~~~~~~~~~~~~~~\mu^\sigma~~~~~~~~~~~~~~~~~$ \\
                         &                         &                              \\
\hline \hline
1.34 & 0.33526(01368)     &              ----              \\
\hline
1.35 & 0.16618(00568)   &                ----   \\
\hline
1.3508  &  0.09530(00400)   &             ----   \\  
\hline
1.3525 &  0.09327(00999)     &              ----              \\
\hline
1.355 & 0.07505(00930)     &              ----              \\
\hline
1.36 & 0.06538(00813)    &              ----              \\
\hline
1.41 &  0.03244(00472)     &            0.20908(00601)           \\
\hline
1.44 & 0.02637(00398)     &              0.21849(00544)              \\
\hline
1.45 &  0.03222(01499)    &            0.19063(01356)          \\
\hline
1.47 & 0.02523(00419)     &              0.15654(00393)             \\
\hline
1.5 & 0.02219(00720)     &              0.18009(00706)             \\
\hline
1.54 & 0.01992(00621)     &             0.16590(00583)              \\
\hline
1.57 & 0.02054(00390)     &              0.12904(00285)              \\
\hline
1.59 & 0.02165(00452)     &              0.11620(00323)              \\
\hline
\end{tabular} 
\label{tabmas} 
\end{table}

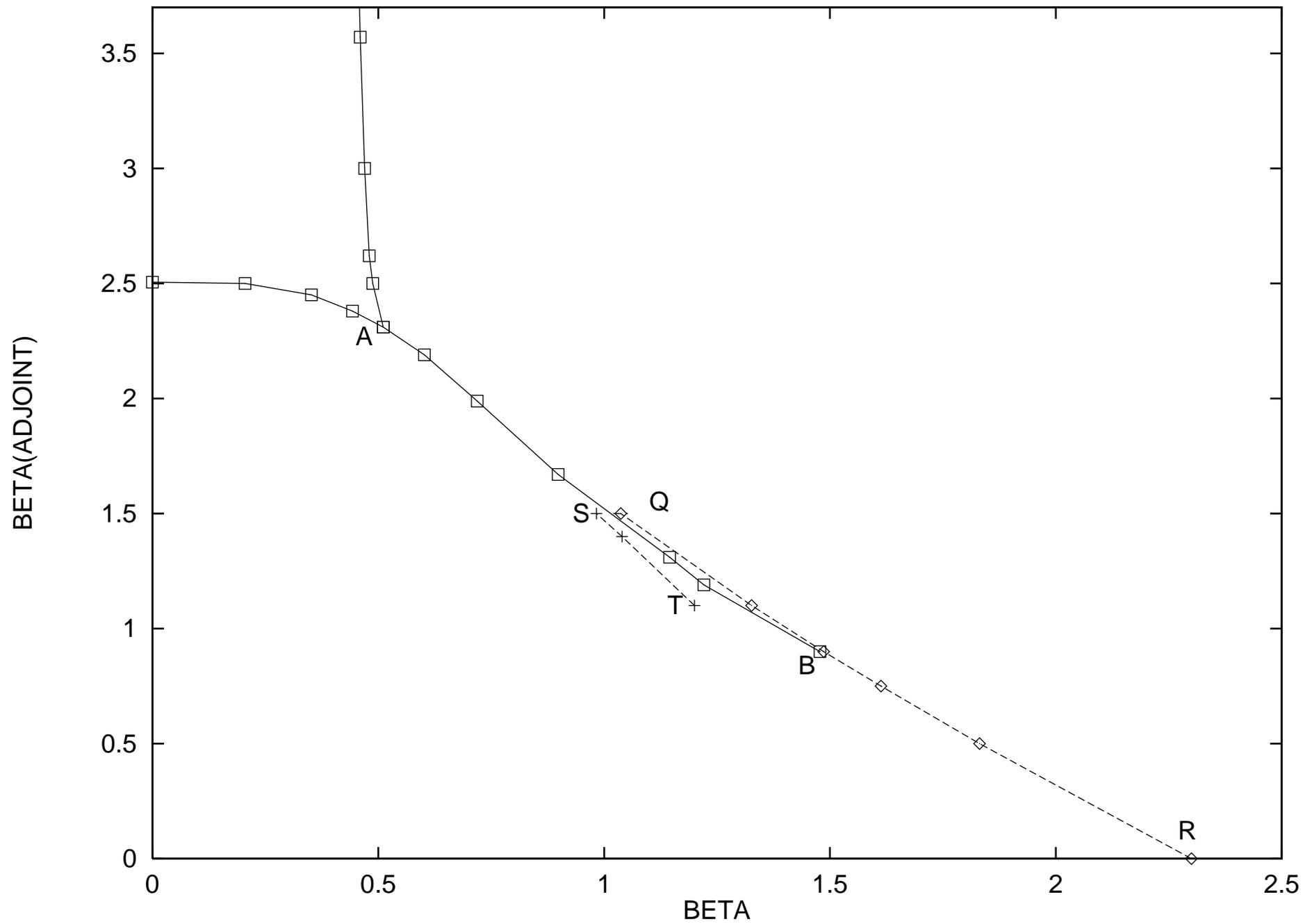
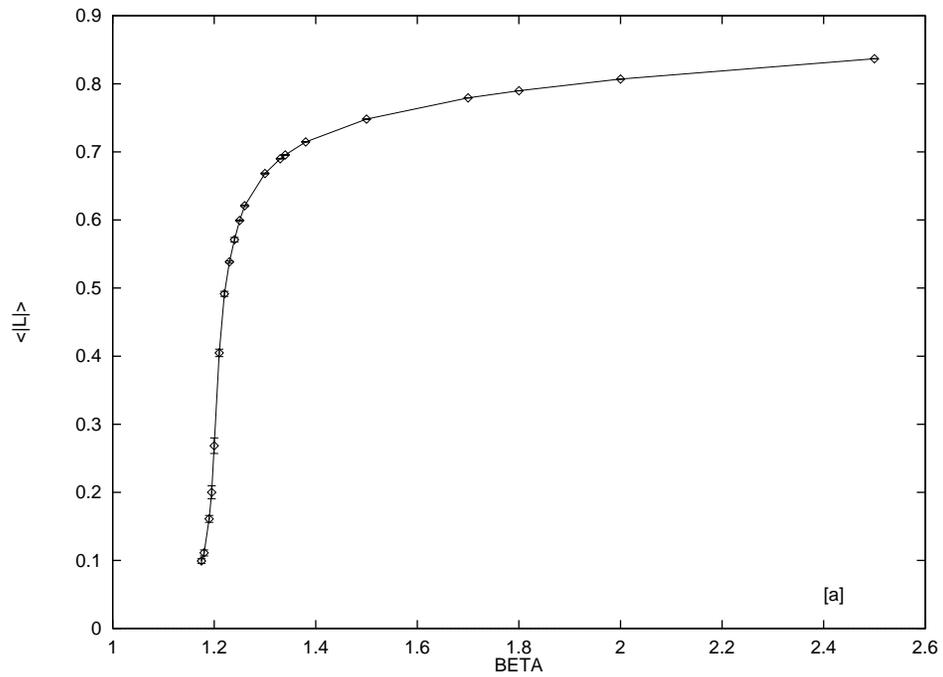
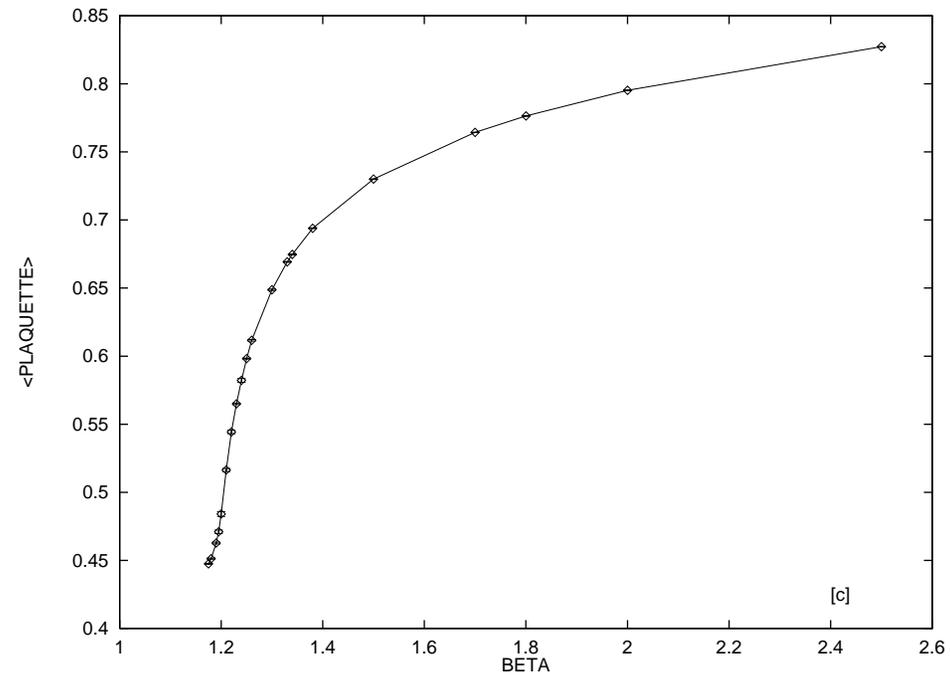
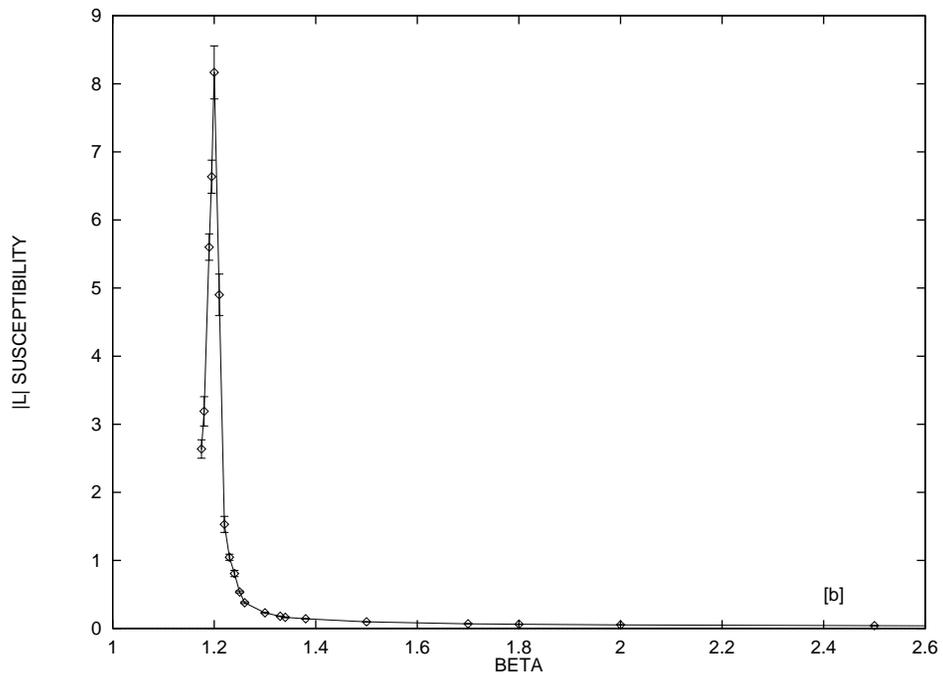
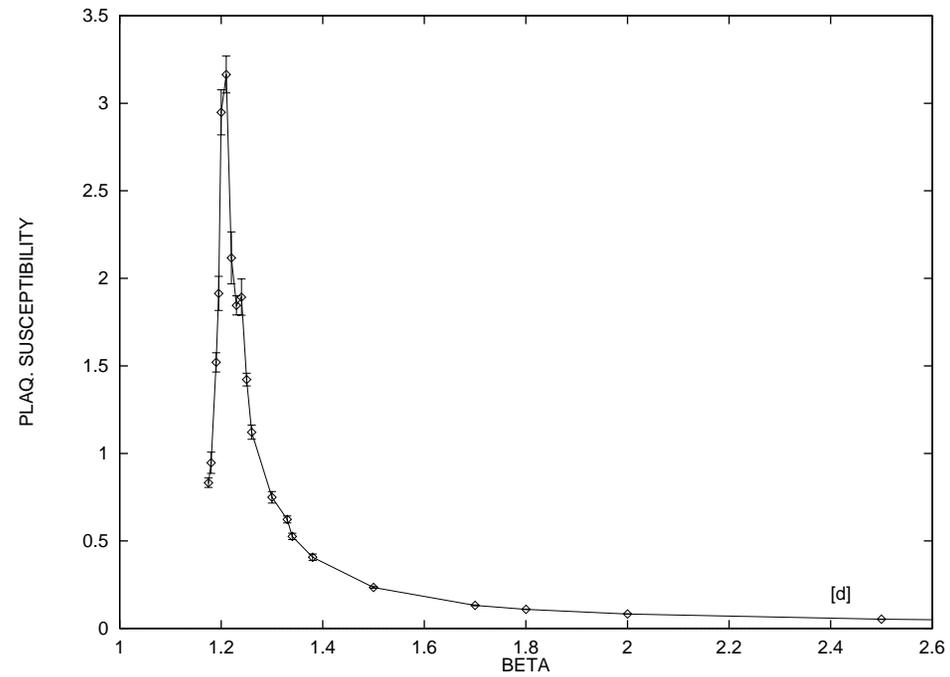
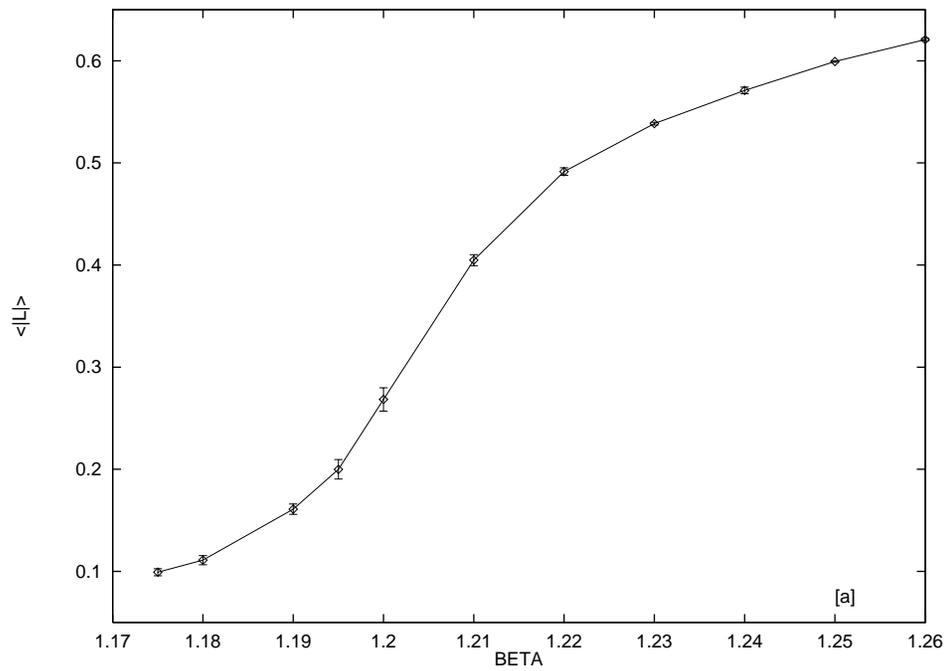
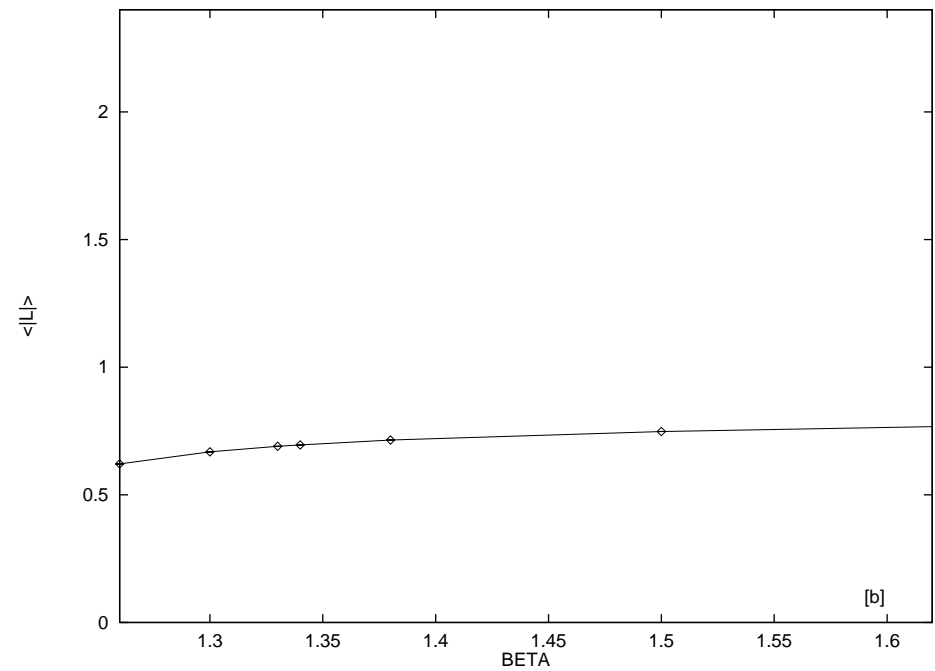
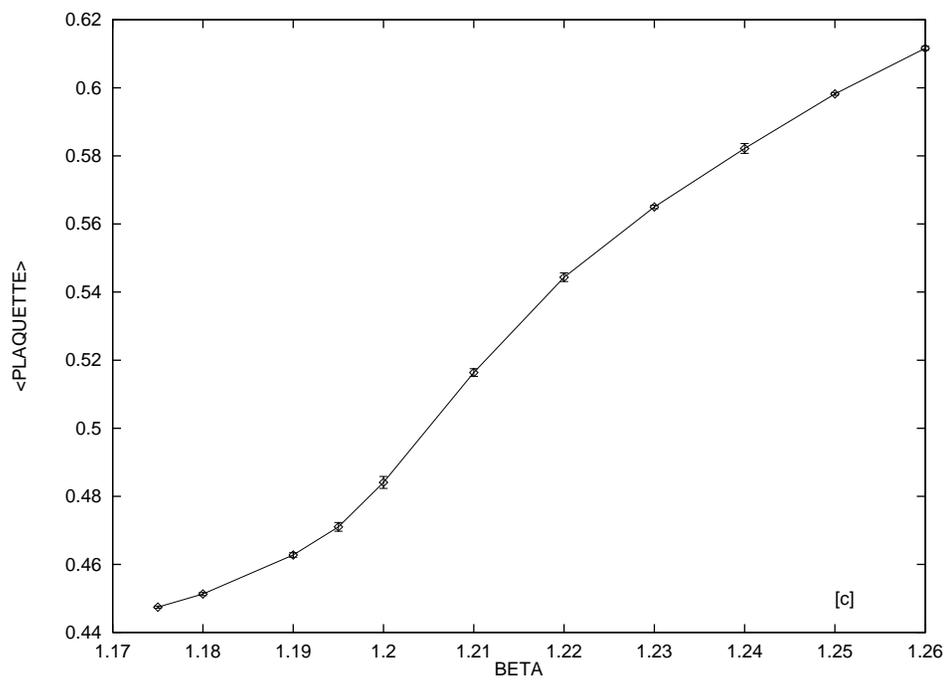
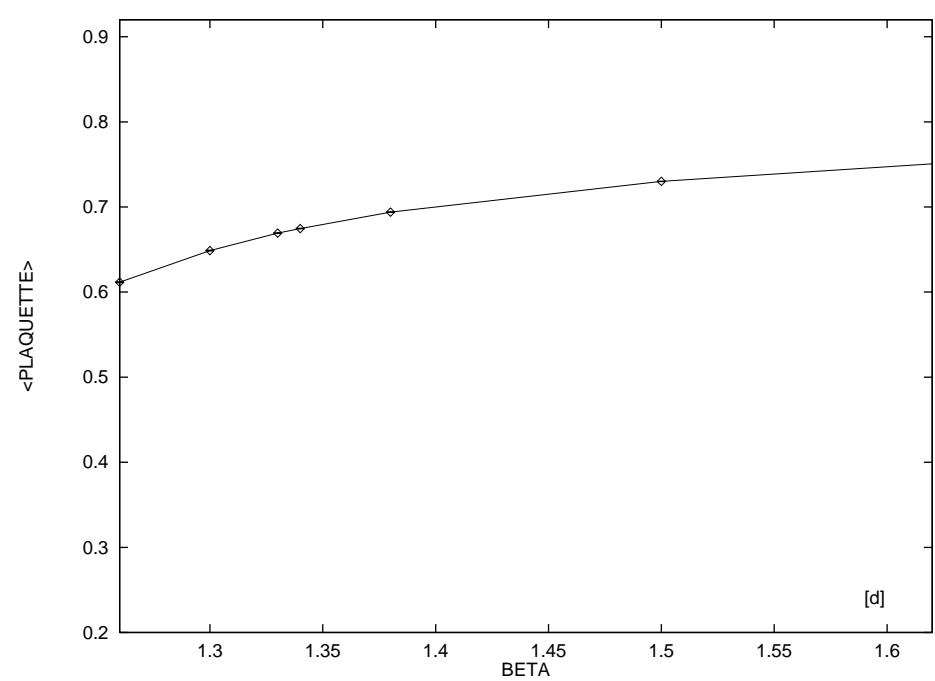
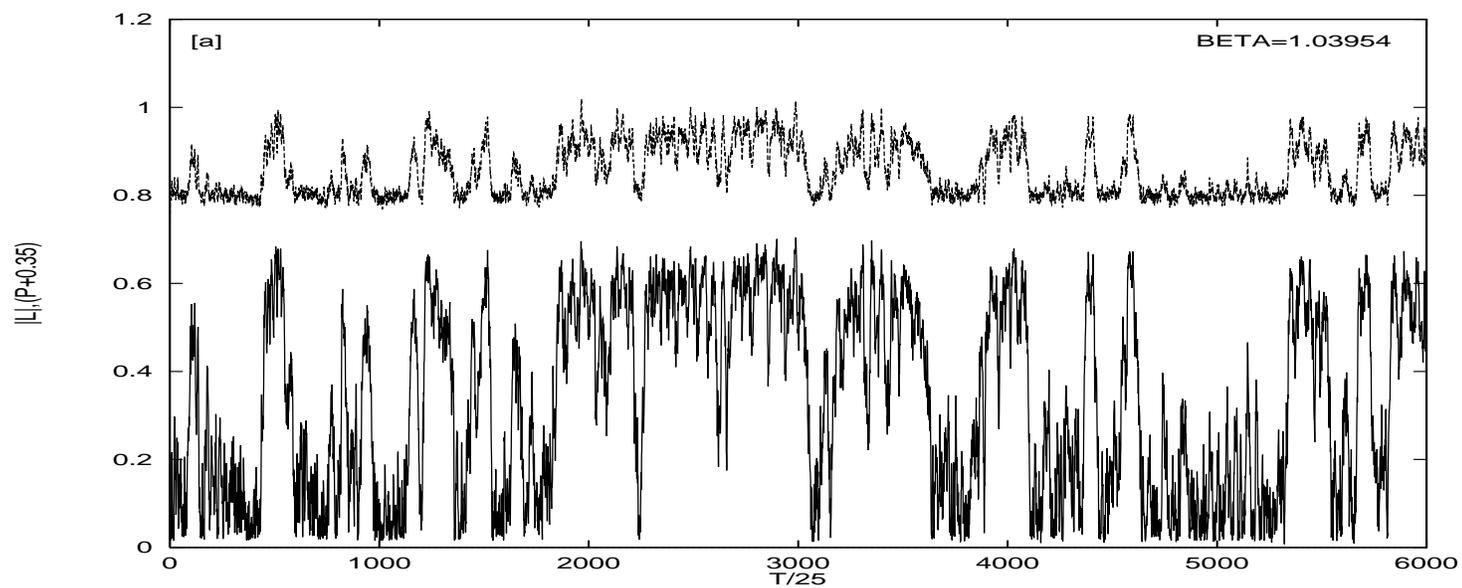
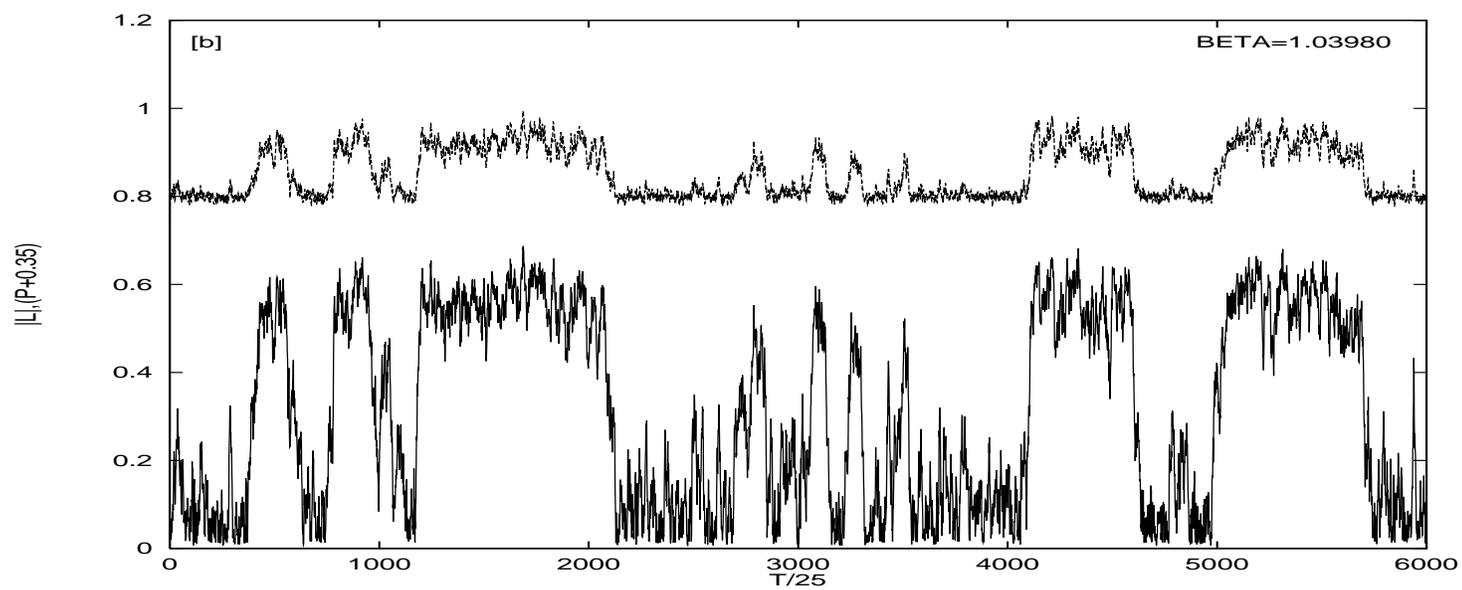
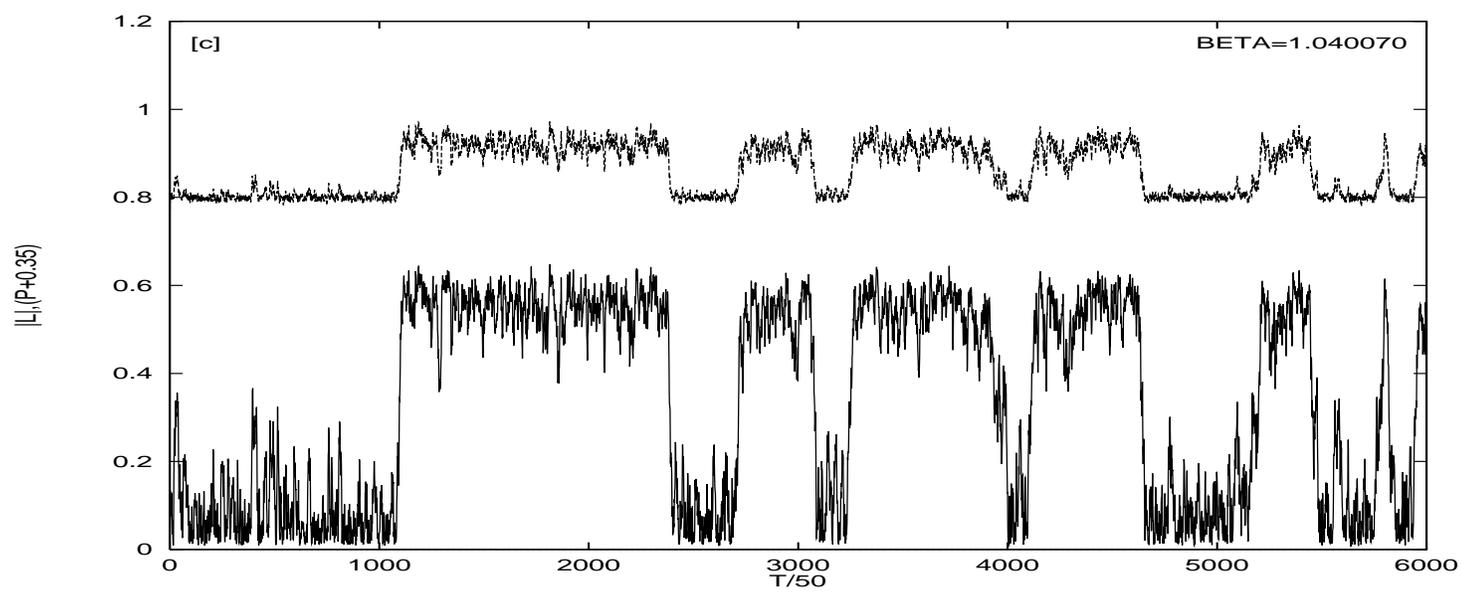
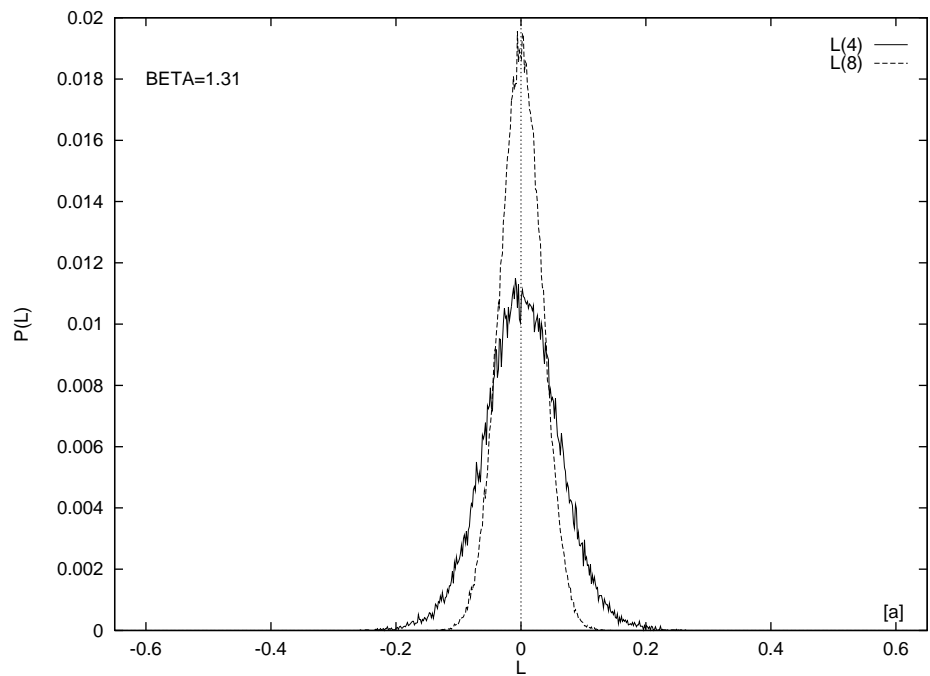
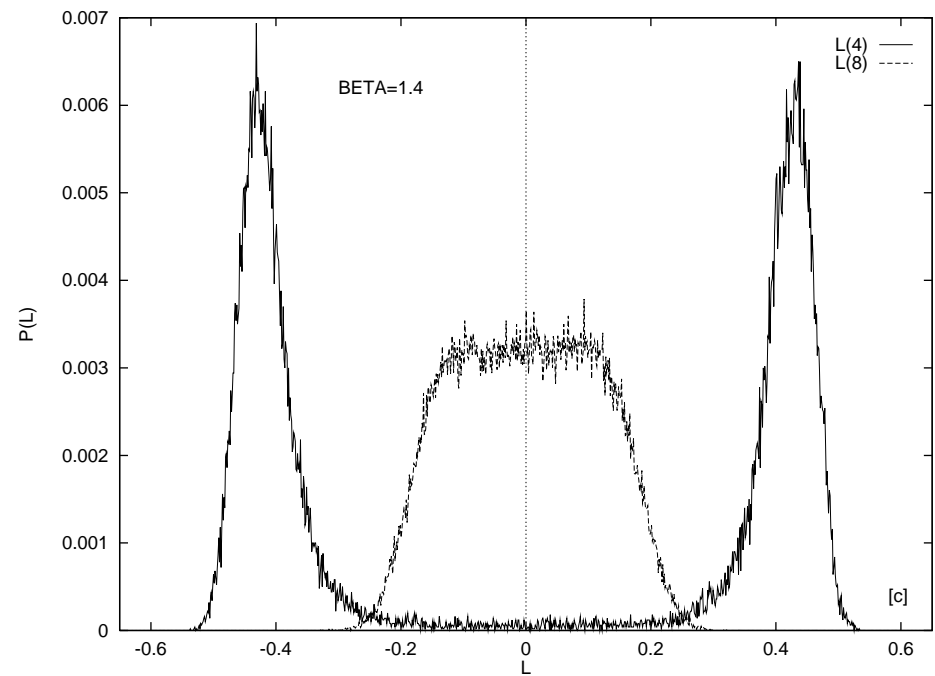
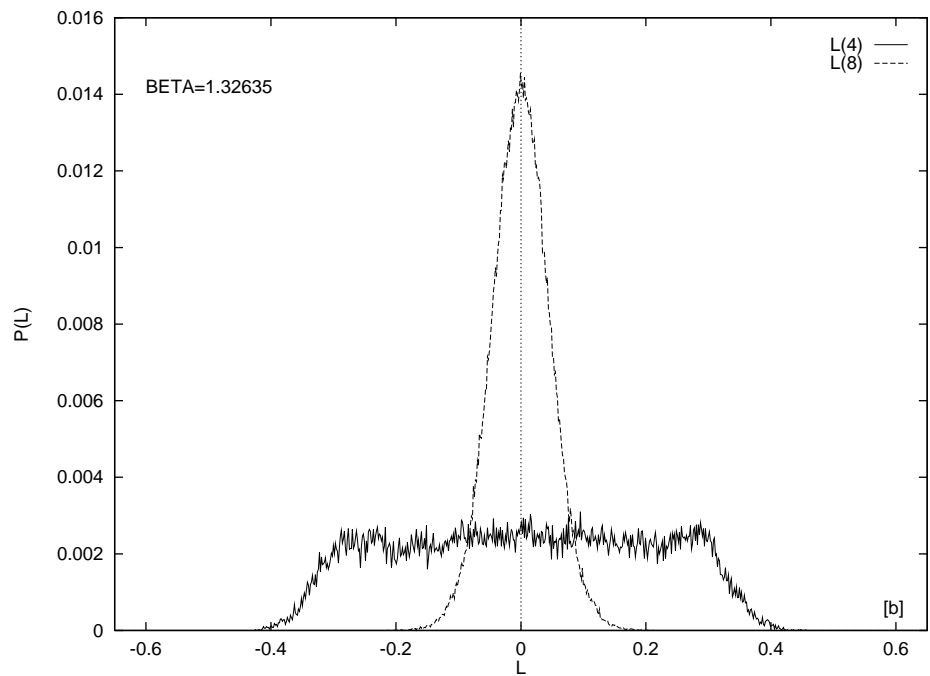
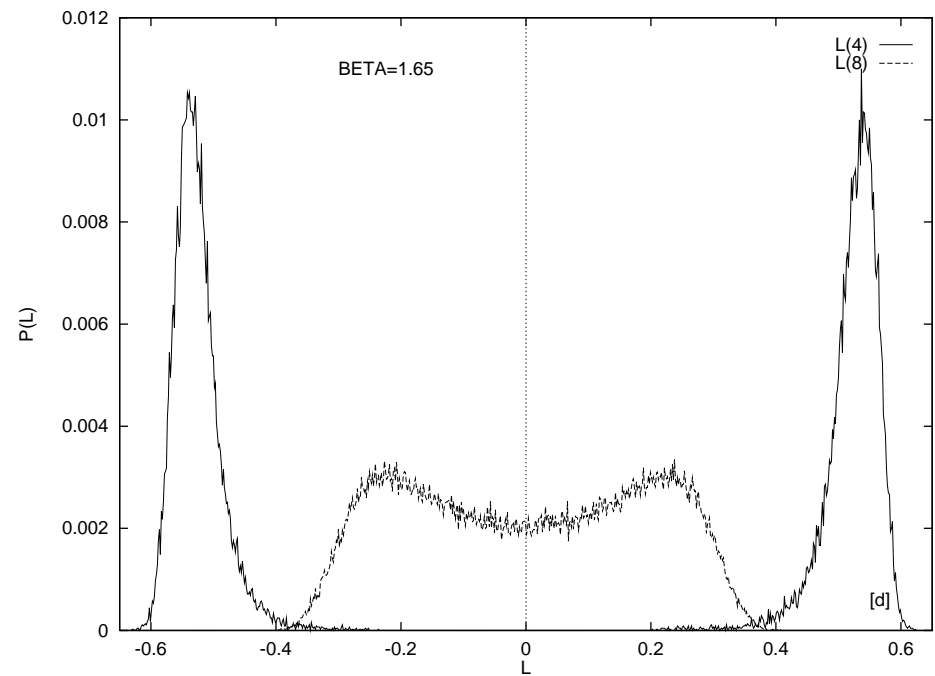
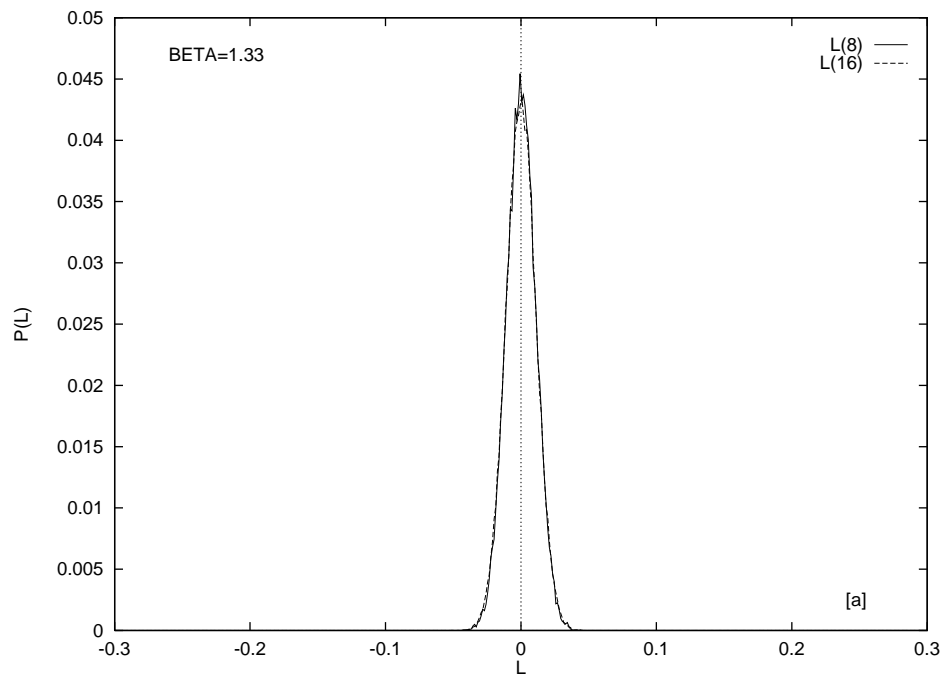
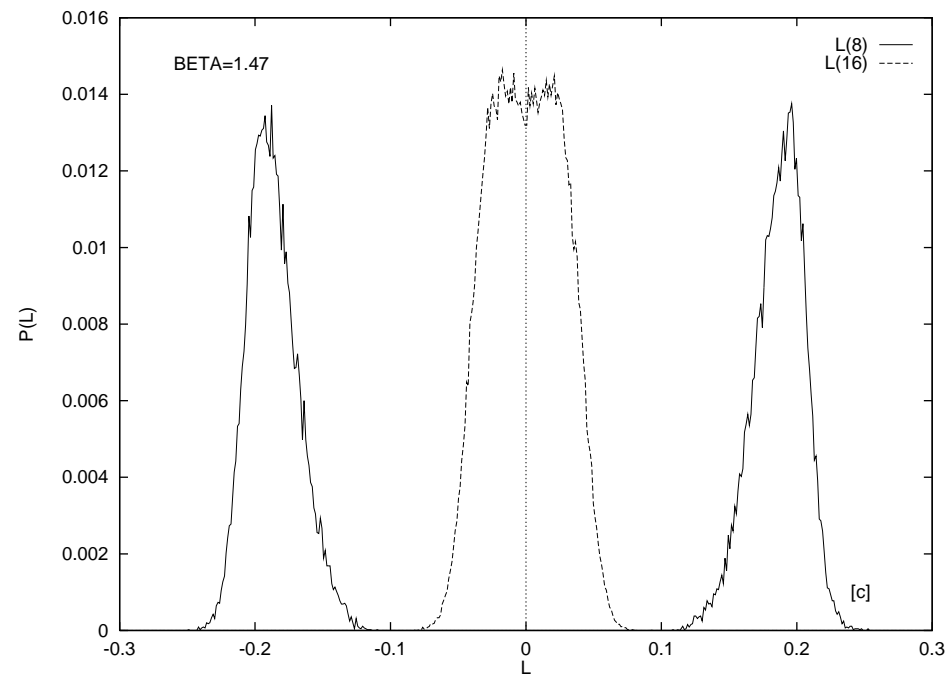
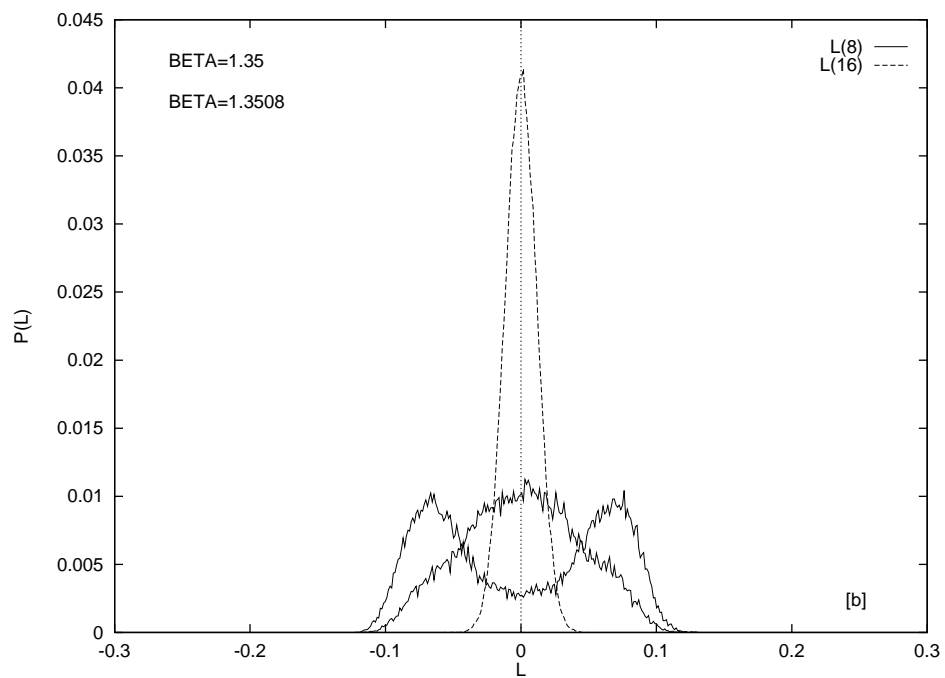
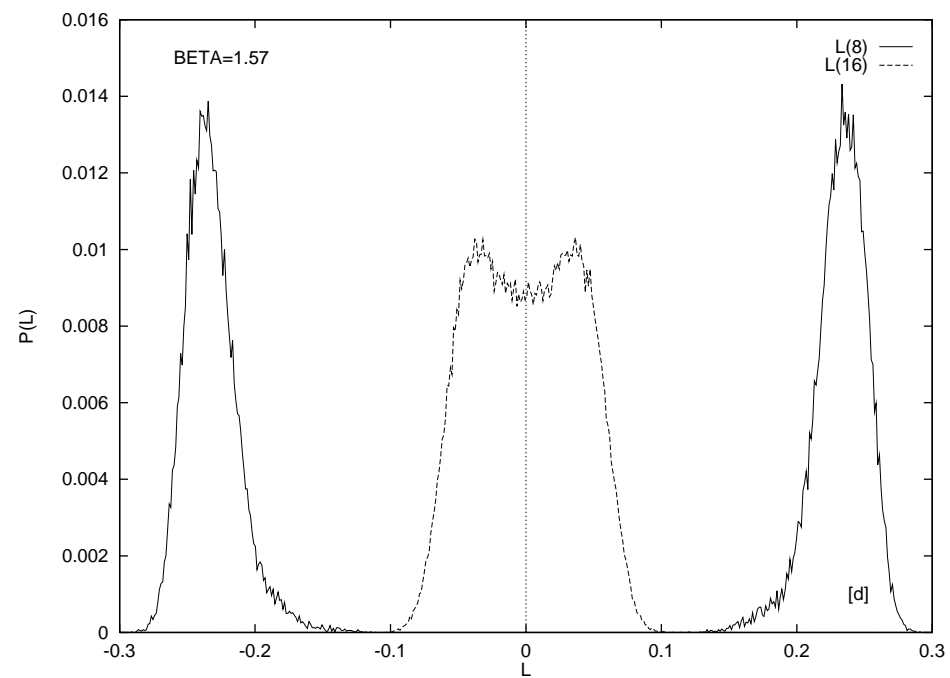
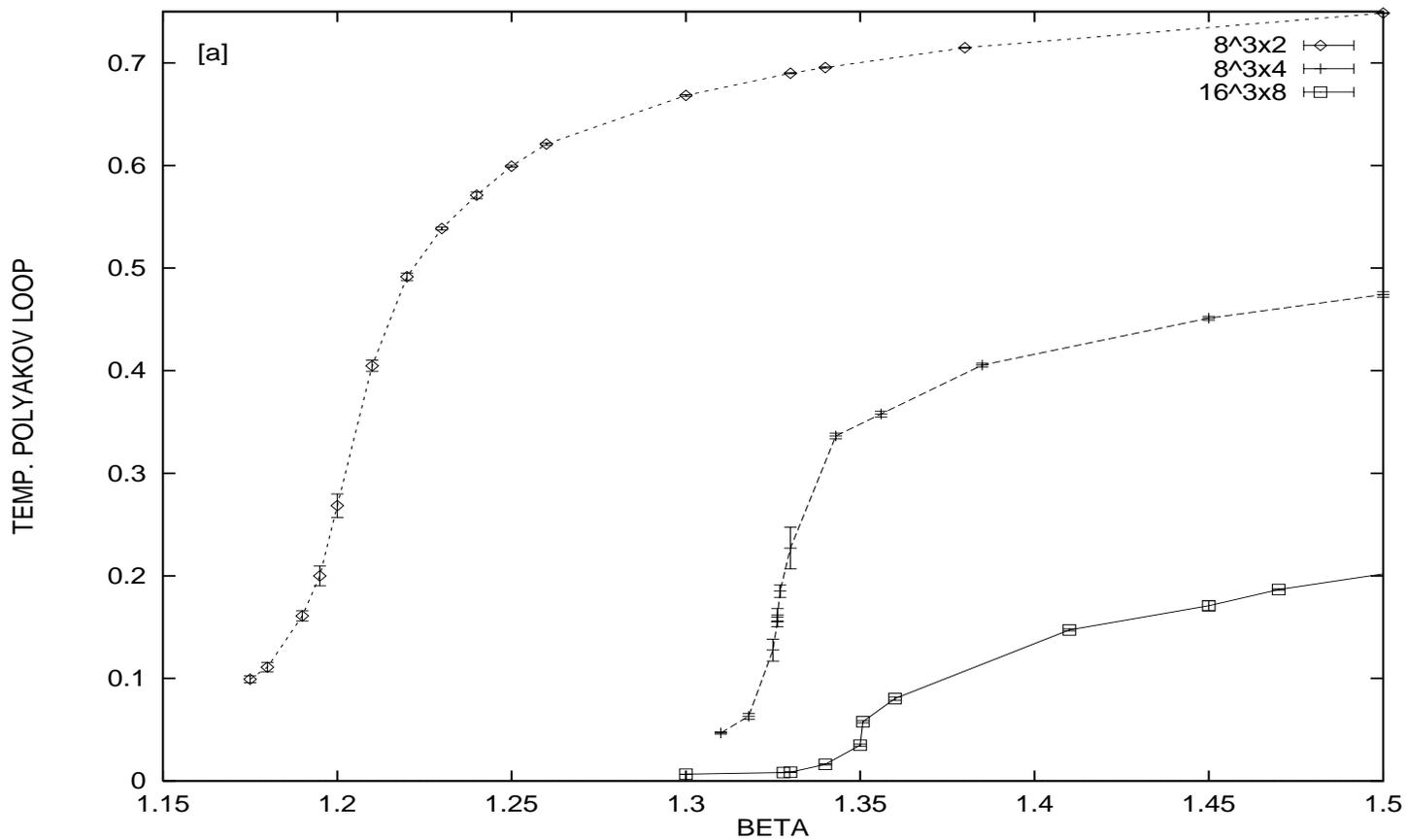
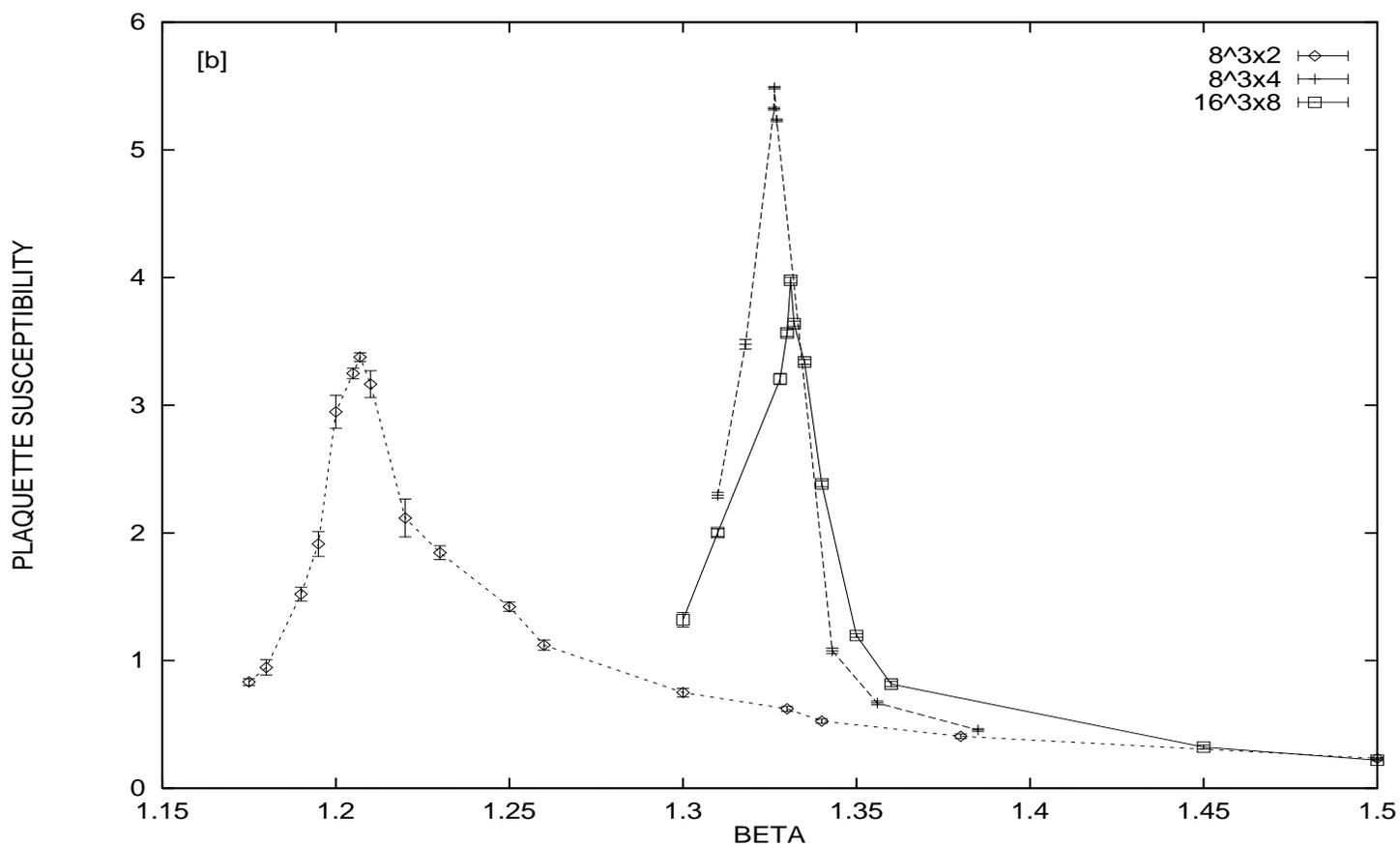
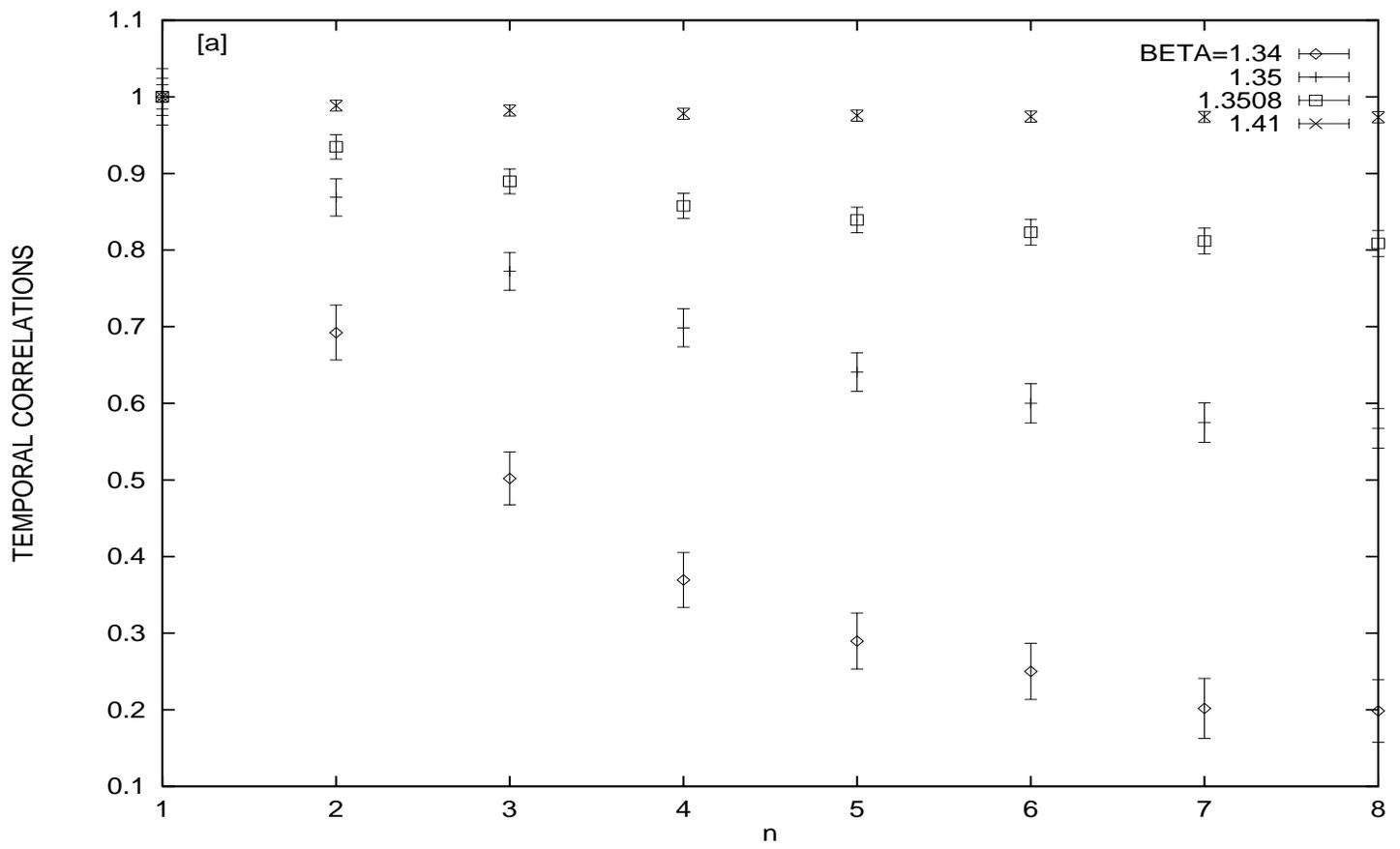
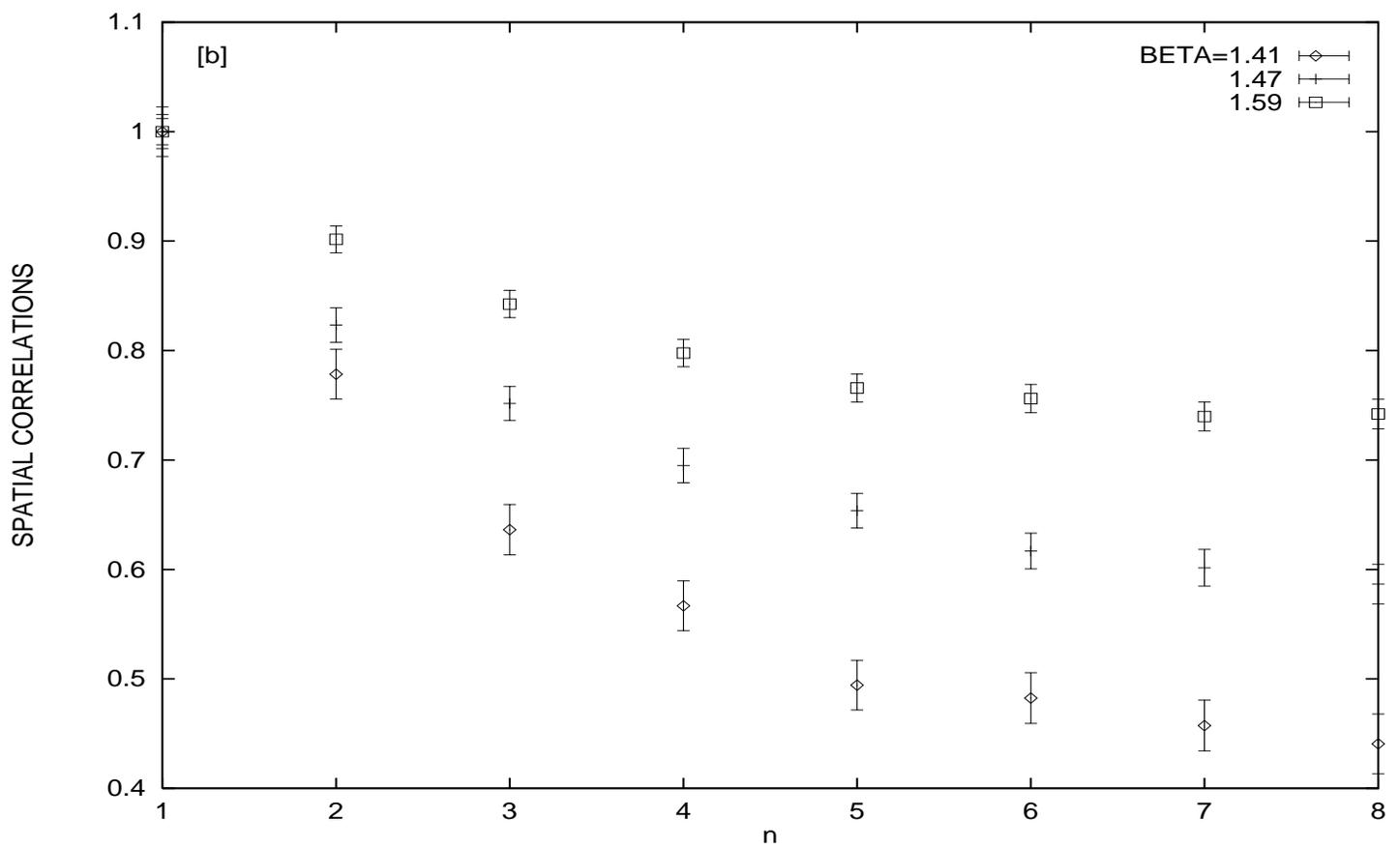
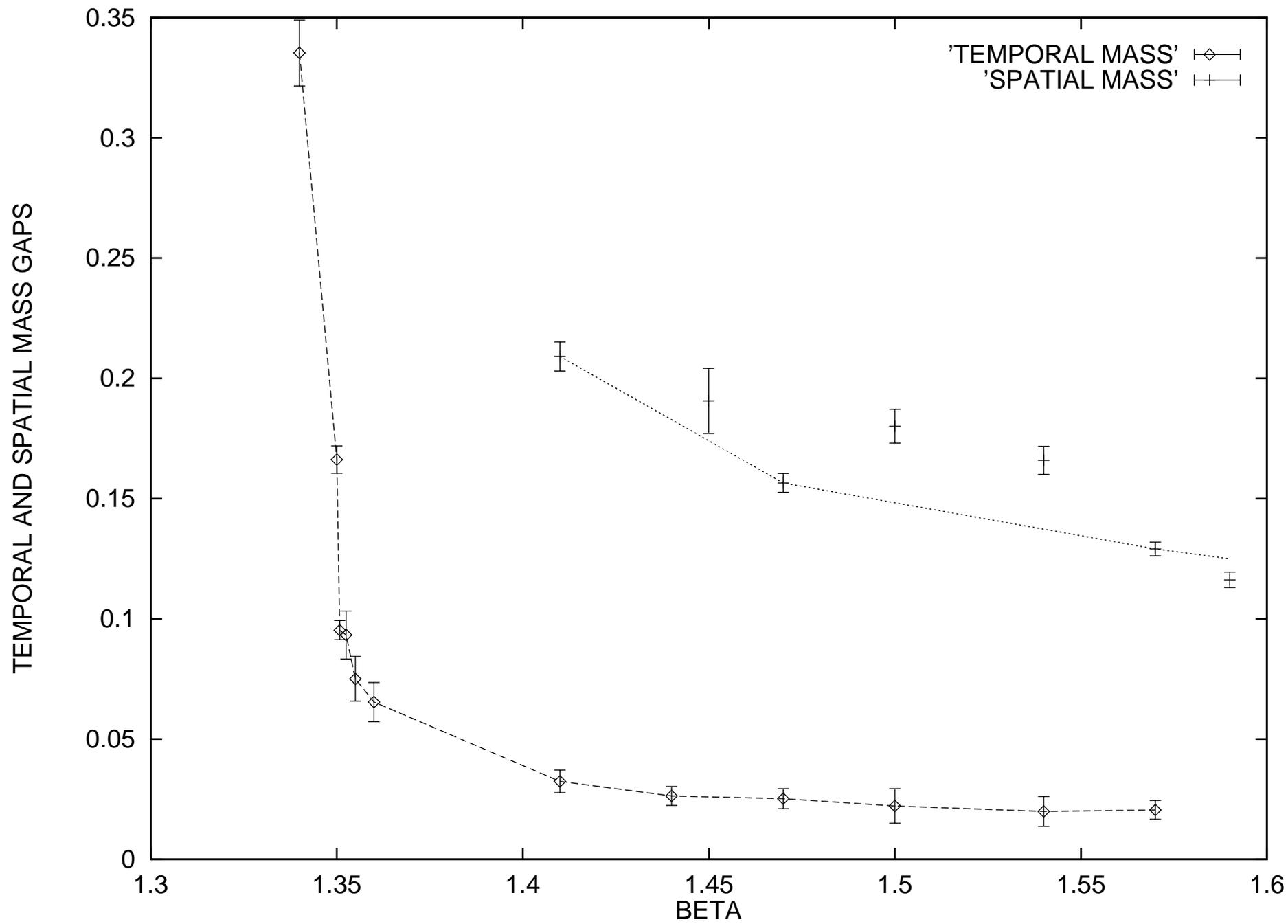
\end{document}